\documentclass[aps,prl,superscriptaddress,amsmath,amssymb,floatfix,twocolumn]{revtex4}
\usepackage{times}
\usepackage{bbold}
\usepackage{graphicx}
\usepackage{subfigure}
\usepackage{color}
\usepackage{hyperref}

\begin{document}

\title{Quantum criticality between topological and band insulators in $(3+1)$-dimensions}
\author{Pallab Goswami}
\affiliation{Department of Physics and Astronomy, Rice University, Houston, TX 77005}
\author{Sudip Chakravarty}
\affiliation{Department of Physics and Astronomy, University of
California Los Angeles\\ Los Angeles, CA 90095-1547}

\date{\today}

\begin{abstract}
Four-component massive and massless Dirac fermions in the presence of long range Coulomb interaction and chemical potential disorder exhibit striking fermionic quantum criticality. For an odd number of flavors of Dirac fermions, the sign of the Dirac mass distinguishes the topological and the trivial band insulator phases, and the gapless semi-metallic phase corresponds to the quantum critical point that separates the  two. Up to a critical strength of disorder, the semi-metallic phase remains stable, and the universality class of the direct phase transition between two insulating phases is unchanged. Beyond the critical strength of disorder the semi-metallic phase undergoes a phase transition into a disorder controlled diffusive metallic phase, and there is no longer a direct phase transition between the two types of insulating phases. Our results are also applicable to even number of flavors of Dirac fermions, and the band inversion transition in various non-topological narrow gap semiconductors.

\end{abstract}


\maketitle

The low energy, long wavelength quasi-particle spectrum of various narrow gap semiconductors is well approximated by  noninteracting $(3+1)$-dimensional massive Dirac fermions and by adjusting the chemical composition, or by applying pressure, the sign of the Dirac mass can be changed at a band-inversion transition (BIT). At BIT the system becomes semi-metallic (SM) and is described by massless Dirac fermions. The SM phase is an interesting example of $z=1$ fermionic quantum critical point (QCP). In the massless phase, the conduction and valence bands cross at a discrete number of diabolic points inside the Brillouin zone and depending on the number of inequivalent diabolic points we obtain multiple flavors of Dirac fermions. The narrow gap semiconductors such as $\textrm{Pb}_{1-x}\textrm{Sn}_{x}\textrm{Te}$ (four flavors), $\mathrm{Bi}_{1-x}\mathrm{Sb}_x$ and $\textrm{Hg}_{1-x}\textrm{Cd}_{x}\textrm{Te}$ (each with single flavor) are well known examples, which for special values of $x$, become massless \cite{Nimtz}. For  odd number of Dirac fermion flavors, the QCP describes the phase transition between a topological insulator (TI) and an ordinary band insulator (BI) \cite{FuKane1,QiZhang1, Zahid1, QiZhang2}. The recent experimental observation of TI phase in different narrow gap semiconductors have spurred our interest in the $(3+1)$-dimensional Dirac materials \cite{Zahid1,QiZhang2}. The low energy spectrum of materials like $\mathrm{Bi}_{1-x}\mathrm{Sb}_x$, $\mathrm{Bi}_2\mathrm{Te}_3$, $\mathrm{Bi}_2\mathrm{Se}_3$, where TI phase has been observed are all described in terms of a single flavor massive Dirac fermion \cite{Liu,Zhang}.

The stability of the disorder free SM phase in the presence of long range Coulomb interaction and the noninteracting SM phase in the presence of random chemical potential disorder were respectively addressed in Refs.~\cite{Abrikosov} and ~\cite{Fradkin}. Recently the noninteracting, disordered $(3+1)$-dimensional TI has been considered in Refs.~\cite{Shindou, Guo}. For the noninteracting disordered problem a symmetry based ten-fold classification of TI and superconductors has been described in Ref.~\cite{Ryu}, and the stability of the two dimensional surface states has been discussed on the basis of this symmetry classification. The noninteracting TI in the presence of generic time reversal symmetric disorder belongs to the symplectic $\mathrm{AII}$ class. In general such classification does not hold in the presence of interaction. The stability of the two dimensional disordered and interacting surface states of a TI has been addressed in Ref.~\cite{Ostrovsky}. However, the combined effects of interaction and disorder on the bulk fermions has not been considered before. Motivated by this and possible future experiments in which novel QCPs can be explored, we analyze the problem of both massive and massless Dirac fermions in the presence of Coulomb interaction and random chemical potential type disorder using a perturbative renormalization group (RG) analysis. Remarkably, the vanishing density of states at the Dirac points renders such a calculation reliable in comparison to the corresponding  non-relativistic problem.

For orientation we first consider the disorder free noninteracting Dirac fermion action. For  simplicity we consider only one species of Dirac fermion. Assuming inversion (parity) and time reversal symmetry and using a spinor basis $\psi^{T}=(c_{+,\uparrow},c_{+,\downarrow},c_{-,\uparrow},c_{-,\downarrow})$, where $c_{\pm,s}$ respectively correspond to the annihilation operators for parity even and odd states, with spin projection $s$, we can write the following Euclidean action
\begin{eqnarray}
S_0=\int d^4x \bar{\psi}\bigg[\gamma_0(\partial_0-A\partial_{j}^{2})+iv \gamma_{j} \partial_{j}+m-B\partial_{j}^{2}\bigg]\psi.
\label{eq:1}
\end{eqnarray}
The latin index $j$ is a spatial index and $\Lambda \sim 1/a$ is the ultraviolet cutoff, where $a$ is the lattice spacing. The parameter  $v$ is the Fermi velocity and $m$ is the Dirac mass. The anticommuting Euclidean $\gamma$ matrices satisfy $\{\gamma_{\mu},\gamma_{\nu}\}=2\delta_{\mu,\nu}$, and $\bar{\psi}=\psi^{\dagger}\gamma_0$. The four-component spinor structure arises from the two-sublattice crystallographic structure and the two spin components. We have also incorporated two higher gradient terms involving $A$ and $B$. The above action is invariant with respect to the parity ($\mathcal{P}$) and time reversal ($\mathcal{T}$) transformations: $\mathcal{P} \psi \mathcal{P}^{-1}=\gamma_0 \psi$, $\mathcal{T} \psi \mathcal{T}^{-1}=-\gamma_1\gamma_3 \psi$. For $A=0$, the action is also invariant under charge conjugation (particle-hole) transformation ($\mathcal{C}$): $\mathcal{C} \psi \mathcal{C}^{-1}=-\gamma_2 \psi$. The particle-hole symmetry breaking term does not affect the topological properties, and can be off-set by adjusting of the chemical potential, and henceforth we will set $A=0$.

The fermion mass $m$ and the higher gradient term Proportional to $B$  break the $U(1)$ chiral symmetry of the massless Dirac fermions defined by $\psi \to e^{i(\theta/2)\gamma_5}\psi$, $\bar{\psi}\to \bar{\psi}e^{i(\theta/2) \gamma_5}$, where $\gamma_5=\gamma_0\gamma_1\gamma_2\gamma_3$. The TI and BI phases are respectively defined by the conditions $mB<0$ and $mB>0$ and   are separated by a finite chiral angle $\delta \theta=\pi$. This is reflected in the quantized magneto-electric coefficients $\pi$ and $0$, respectively for TI and BI. At the critical point $m=0$, the dynamic exponent $z=1$, and in the RG sense the higher derivative terms can be ignored.

The Dirac structure of the Hamiltonian allows various types of disorder. The constraint of time reversal invariance allows the following six bilinears:  $\bar{\psi}\gamma_0\psi$, $\bar{\psi}\psi$, $\bar{\psi}\gamma_0\gamma_5\psi$ and $\bar{\psi}\gamma_0 \gamma_{j} \psi$ ($j=1,2,3$). The bilinears $\bar{\psi}\gamma_0\psi$, $\bar{\psi}\psi$ respectively correspond to random chemical potential and random mass scattering. The physical description of other four bilinears depends on the crystallographic details. We shall concentrate on the random chemical potential as the dominant elastic scattering process, and add $S_D=\int d^4x V(\mathbf{x})\bar{\psi}\gamma_0 \psi $ to the action $S_0$. The random potential $V(\mathbf{x})$ is a Gaussian white noise distribution specified by the disorder average $\langle\langle{V(\mathbf{x})V(\mathbf{x}^{'})}\rangle\rangle=\Delta_V\delta^{3}(\mathbf{x}-\mathbf{x}^{'})$. The detailed analysis for generic time reversal symmetric disorder is provided in the supplementary material \cite{Supp}.

Since typically $v/c \sim 10^{-2}-10^{-3}$ ($c$ is the velocity of light), the Coulomb interaction is instantaneous. Its strength  is characterized by the the dimensionless parameter $\alpha =e^2/(4\pi \varepsilon v)\sim 2.2/\varepsilon - 22/\varepsilon$, where $\varepsilon$ is the static dielectric constant of the material. We perform disorder average using the replica method, which we use merely as a book-keeping device for perturbative RG calculations. The  replicated  Euclidean action  after disorder averaging of the partition function is
\begin{eqnarray}
\overline{S}=\int d^4x \bigg[\bar{\psi}_{a}\left\{\gamma_0(\partial_0+ig\varphi_{a})+v\gamma_{j}\partial_{j}+m -B\partial_{j}^{2}\right\}\psi_{a}\nonumber \\+\frac{1}{2}(\partial_{j} \varphi_{a})^{2}\bigg]
-\frac{\Delta_{V}}{2}\int d^3x dx_0 dx_0^{'}\left(\bar{\psi}_{a}\gamma_{0}\psi_{a}\right)_{(x,x_0)} \nonumber \\ \times \left(\bar{\psi}_{b}\gamma_{0}\psi_{b}\right)_{(x,x_{0}^{'})}
\label{eq:2}
\end{eqnarray}
where $g=\sqrt{4\pi v\alpha}$, and $a$, $b$ are replica indices. We have used the abbreviated notation $\left(\bar{\psi}_{a}\gamma_{0}\psi_{a}\right)_{(x,x_0)}\equiv \bar{\psi}_{a}(x,x_{0})\gamma_{0}\psi_{a}(x,x_{0})$. We have introduced an auxiliary scalar potential $\varphi_{a}$ to decouple the four-fermion Coulomb interaction term. Unlike two dimensions, the Coulomb interaction manifests itself as $(\partial_{j} \varphi_{a})^{2}$, which is analytic in momentum and $g$ can certainly receive loop corrections. The action $\overline{S}$ preserves all three discrete symmetries $\mathcal{P}$, $\mathcal{T}$ and $\mathcal{C}$. For RG calculations we replace the couplings by the corresponding dimensionless couplings $m \to m/(v\Lambda)$, $B \to B\Lambda /v$ and $\Delta_V \to \Delta_V \Lambda /(2\pi^2v^2)$. The details of the RG calculation are provided in the supplementary material~\cite{Supp}.

The density of states for massless and massive problems are,  respectively, $\rho (E) \propto E^2$ and $\rho (E) \propto |E|\sqrt{E^2-m^2}$. Since the the density of states vanishes at zero energy, the scattering rate $\tau^{-1}(E)$ calculated from lowest order Born approximation also vanishes at zero energy. At the tree level, $B$ and $\Delta_V$ are irrelevant couplings, and $\alpha$ and $m$ are respectively marginal and relevant couplings. Therefore we anticipate that the universality class of the QCP between TI and BI will be unchanged up to a critical strength of the disorder. This should be contrasted to the two dimensional problem, where the chemical potential disorder is a marginally relevant perturbation and invalidates the lowest order Born approximation result~\cite{Foster}.

\begin{figure}[htb]
\centering
\subfigure[]{
\includegraphics[scale=0.8]{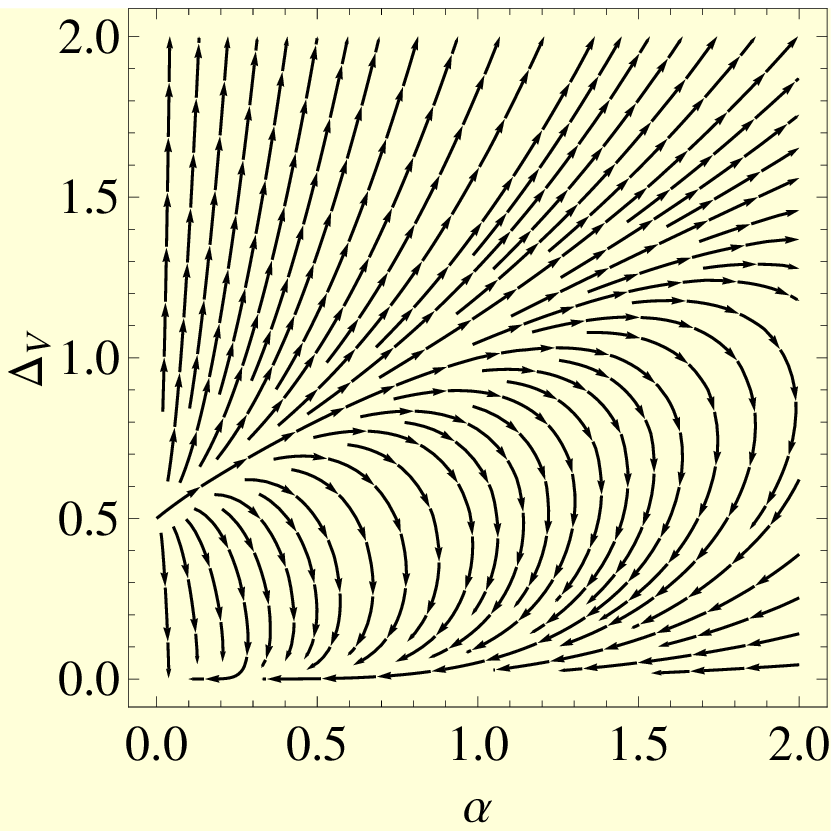}
\label{fig:subfig1a}
}
\subfigure[]{
\includegraphics[scale=0.8]{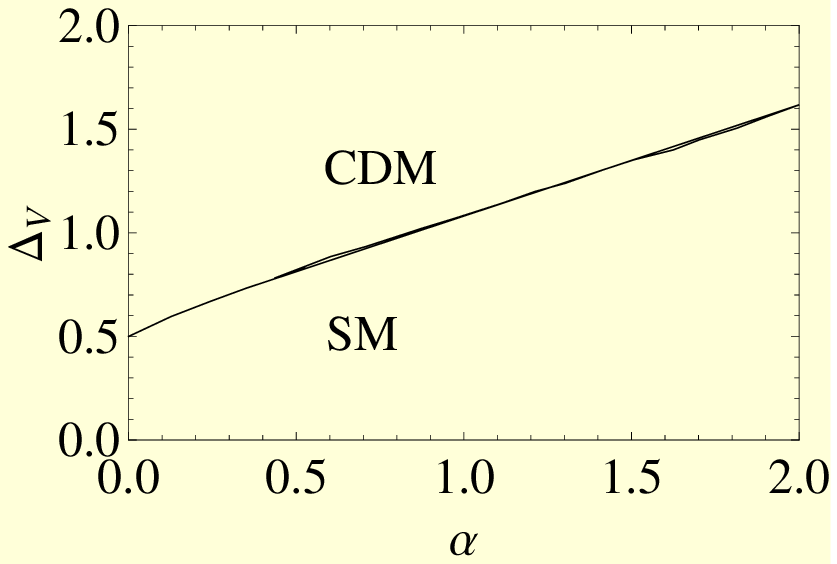}
\label{fig:subfig1b}
}
\label{fig:1}
\caption[]{(a)The RG flow and (b) the phase diagram in the $\alpha-\Delta_V$ plane for massless Dirac fermions with $m=B=0$. The CDM phase represents a disorder induced massless, compressible diffusive metallic state with finite density of states and scattering rate at zero energy.
}
\end{figure}
 Consider first the massless Dirac fermions with $B=0$. From a momentum shell renormalization group (RG) calculation to ${\cal O}(\alpha^2)$, ${\cal O}(\Delta_{V}^{2})$, and ${\cal O}(\Delta_V\alpha)$, we find
\begin{eqnarray}
&&\frac{dv}{dl}=v\left(z-1-\Delta_V+\frac{2\alpha}{3\pi}\right),\\
&&\frac{d \alpha}{dl}=\alpha \left(\Delta_V-\frac{4\alpha}{3\pi}\right),\\
&&\frac{d\Delta_V}{dl}=\Delta_V \left(-1+2\Delta_V-\frac{8\alpha}{3\pi}\right).
\end{eqnarray}
By keeping $v$ fixed we obtain a scale dependent dynamic exponent $z(l)=1+\Delta_V(l)-2\alpha(l)/3\pi$. There are two fixed points: (i) attractive, noninteracting, clean fixed point:  $\Delta_V=\alpha=0$, $z=1$; and (ii)noninteracting finite disorder critical point: $\Delta_V=1/2$, $\alpha=0$ , $z=3/2$. For the noninteracting disordered problem, the fixed point (ii) controls the transition between the two phases where disorder is respectively irrelevant (SM) and relevant. In the phase where disorder is relevant, both the zero energy density of states, and the zero energy scattering rates are finite. Therefore this disorder induced phase will be termed a compressible diffusive metal (CDM).  The RG flow and the associated phase diagram are respectively shown in Fig.~\ref{fig:subfig1a} and Fig.~\ref{fig:subfig1b}.

By linearizing the flow equations in the vicinity of the fixed point (ii), we find $\alpha(l)\approx \alpha_0e^{l/2}$, and $\Delta_V-1/2-8\alpha/(3\pi)\approx (\Delta_{V0}-1/2-8\alpha_{0}/(3\pi))e^{l}$. Therefore at the disorder controlled critical point $\alpha$ is a relevant perturbation, and $\alpha$ shifts the SM-CDM phase boundary to a larger value of $\Delta_V=1/2+8\alpha/(3\pi)$. Along this phase boundary the correlation length diverges with an exponent $\nu=1$, but the dynamic exponent varies continuously as $z=3/2+2\alpha/\pi$. Notice that CDM phase is a strongly interacting state of matter, and this was not addressed in Ref.~\onlinecite{Fradkin,Shindou,Guo}. In the SM phase, the Coulomb interaction initially grows before curling back towards zero in a logarithmic manner, and the initial growth of $\alpha$ is controlled by the bare strength of $\Delta_V$. This unusual  flow will be reflected as a non-monotonic temperature dependence of the inelastic scattering rate. The critical behavior at the SM-CDM phase boundary should be contrasted with its $(2+1)$-dimensional counterpart. In $(2+1)$-dimensions, there is no perturbative loop correction to $g$, and the phase boundary is a line of critical points with $z=1$\cite{Foster}.

\begin{figure}[ht]
\centering
\subfigure[]{
\includegraphics[scale=0.8]{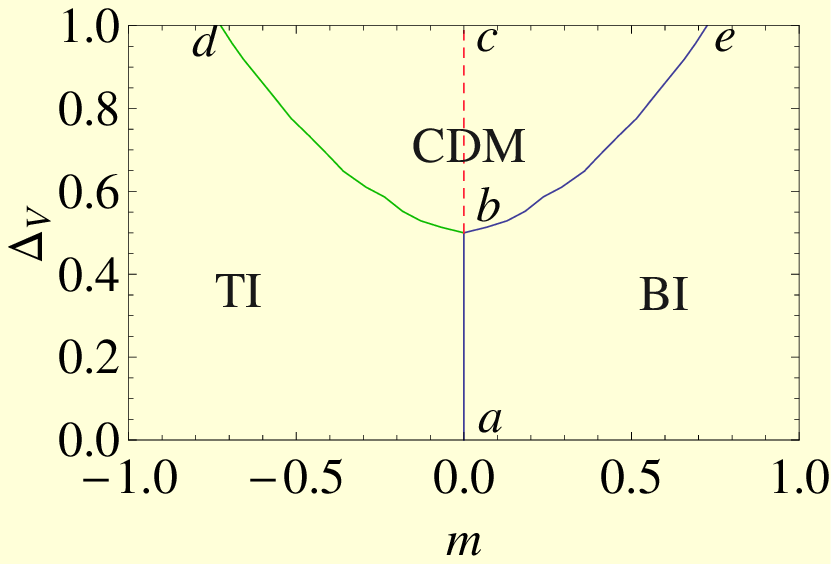}
\label{fig:subfig2a}
}
\subfigure[]{
\includegraphics[scale=0.8]{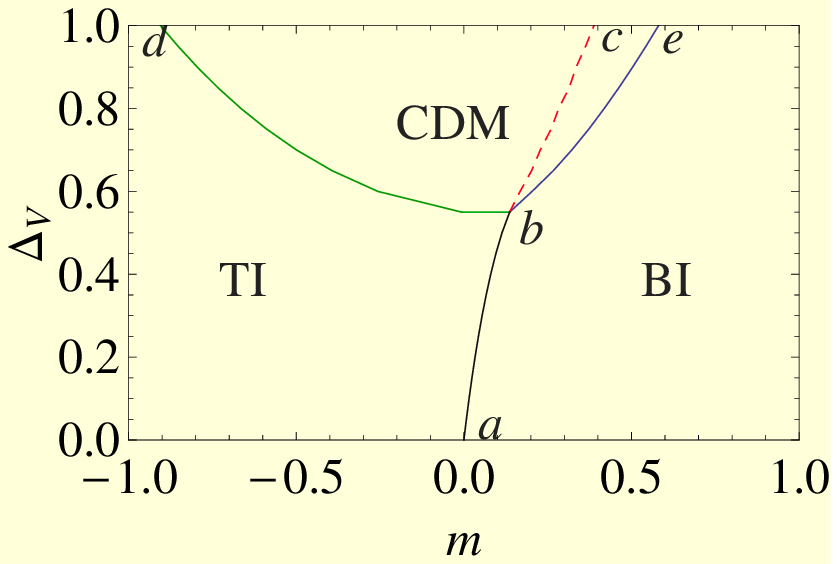}
\label{fig:subfig2b}
}
\label{fig:2}
\caption[]{The phase diagrams in the $m-\Delta_V$ plane for (a)$B_0=\alpha_0=0$, (b)$B_0=0.5$, $\alpha=0$. The direct transition between TI and BI along $ab$ is governed by massless Dirac fermions. When disorder exceeds a critical strength, the insulating phases are separated by CDM, and transitions along $bd$, and $be$ have non-universal critical properties. The dashed line $bc$ describes a cross-over between two regions of CDM with negative and positive effective masses.
}
\end{figure}

Now we consider the role of $m$ and $B$. Compared to the SM phase, we expect TI and BI to be stable up to a larger  disorder, $\Delta_V(m)> (1/2+8\alpha/(3\pi))$. Due to the irrelevant nature of $B$, we expect it to cause non-universal shift of the phase boundaries, leaving the critical properties unchanged. For finite $m$ and $B$, the RG equations are given by
\begin{widetext}
\begin{eqnarray}
&&\frac{dv}{dl}=v\left[z-1+\frac{2\alpha}{3\pi \sqrt{1+(m+B)^2}}-\frac{\Delta_V}{1+(m+B)^2}\right]\\
&&\frac{dm}{dl}=m\left[1+\frac{\alpha}{3\pi \sqrt{1+(m+B)^2}}-\frac{\Delta_V}{1+(m+B)^2}\right]+B\left[\frac{\alpha}{\pi \sqrt{1+(m+B)^2}} -\frac{\Delta_V}{1+(m+B)^2}\right]\\
&&\frac{dB}{dl}=-B\left[1+\frac{\alpha}{3\pi \sqrt{1+(m+B)^2}}\right]+m\frac{\alpha}{3\pi \sqrt{1+(m+B)^2}}\\
&&\frac{d\alpha}{dl}=\alpha \left[\frac{\Delta_V}{1+(m+B)^2}-\frac{2\alpha}{3\pi \sqrt{1+(m+B)^2}}-\frac{2\alpha }{3\pi}\frac{1+\frac{3}{2}(m^2+B^2)+mB}{\left[1+(m+B)^2\right]^{\frac{5}{2}}}\right]\\
&&\frac{d\Delta_V}{dl}=\Delta_V\left[-1+\frac{2\Delta_V}{1+(m+B)^2}-\frac{4\alpha}{3\pi \sqrt{1+(m+B)^2}}-\frac{4\alpha }{3\pi}\frac{1+\frac{3}{2}(m^2+B^2)+mB}{\left[1+(m+B)^2\right]^{\frac{5}{2}}}\right]
\end{eqnarray}
\end{widetext}
In Fig.~\ref{fig:subfig2a} we show the phase diagram for $B=\alpha=0$ and in Fig.~\ref{fig:subfig2b} we show the phase diagram for $\alpha=0$ and a bare value $B_0=0.5$.
When disorder is irrelevant, there is a direct phase transition between TI and BI phases along the line $ab$. In this region, $\Delta_V(l) \sim \Delta_V e^{-l}$, and $B(l) \sim B_0e^{-l}$, and $\alpha(l) \sim \alpha_0(1+4\alpha_0l/(3\pi))^{-1}$, and only relevant variable is fermion mass $m$. In this region for $\alpha_0=0$, we find
$(m(l)-\Delta_V(l)B_(l)/3)\approx e^l(m_0-\Delta_{V0}B_0/3)$. Therefore $(m-\Delta_VB/3)$ behaves as the effective mass, and for finite $B$, the TI-BI phase boundary shifts to $m=B\Delta_V/3$. This can be seen by comparing the segment $ab$ in Fig.~\ref{fig:subfig2a} and Fig.~\ref{fig:subfig2b}. The Coulomb interaction causes additional shift to $m=B\Delta_V/3+B\alpha/(2\pi)$. Therefore in the weak disorder regime, it is possible to induce a transition between two insulating phases by tuning the strength of the disorder. As $\Delta_V$ and $\alpha$ are respectively irrelevant and marginally irrelevant couplings, $z$ asymptotically approaches unity, and the universality class is described by the massless Dirac fermions. There are logarithmic corrections to the scaling properties due to marginally irrelevant nature of $\alpha$ and it is captured through the scale dependent $z$, and also by a factor $e^{l}(\alpha_0/\alpha)^{1/4}$ for the scaling dimension of $m$. The point $b$ is a multi-critical point at which the massless SM phase undergoes a transition either into CDM or one of the two insulating phases. When disorder exceeds the critical strength corresponding to $b$, there is no longer a direct transition between two insulating phases. Along the TI-CDM and BI-CDM phase-boundaries respectively denoted by $bd$ and $be$, $z$ is non-universal, but the mean free path still has the exponent unity. The dashed line $bc$ separates the disorder controlled CDM phase into two regions with negative and positive effective masses, which do not have any physical distinction. For a special case of chiral symmetric disorder and $B=0$, the chiral symmetric diffusive metallic phase along $bc$ becomes distinct from the rest of the CDM phase. In that case, due to the presence of additional diffusive modes, the weak anti-localization correction for a chiral symmetric CDM becomes two times larger than that of the chiral symmetry breaking CDM \cite{Shindou}. However for $B\neq0$, or generic disorder such distinction is lost and $bc$ corresponds to a cross-over line. The Coulomb interaction shifts the point $b$ to higher strength of disorder, and leads to additional non-universal shifts of the phase boundaries, and non-universal change of $z$.

In the weak disorder regime, the transition between TI and BI will be accompanied by interesting critical properties of massless Dirac fermions \cite{Supp}. Since $(3+1)$-dimensions is marginal, some care is necessary to disentangle slow logarithmic corrections in many physical quantities. For example, the specific heat $C_V$ and compressibility $\kappa$, instead of being proportional to $T^3/v^3$ will be proportional to $(T^3/v^3)\{1+4\alpha_0/(3\pi)\log(v\Lambda/T)\}^{-3/2}$. Similar logarithmic corrections in $(2+1)$-dimensions have been discussed in Ref.~\cite{Schmalian}. In the high temperature limit, the diamagnetic susceptibility instead of being a constant has logarithmic enhancement $\chi \approx -e^2v/(24\pi^2)\log(v\Lambda/T)$. A similar logarithmic correction proportional to $\log(B)$ appears in the strong field limit. However a finite particle-hole symmetry breaking term $A$ and $\mu$ will lead to conventional diffusive Fermi liquid behavior in the low temperature limit specified by $\mu/T\gg1$.  Therefore the critical behavior will be limited to $T \gg \mu$. However by a careful adjustment of $\mu$, the critical properties can be found even in the low $T$ limit. The critical behavior will be found even for the massive fermions provided that $T>m$. In the critical regime the inelastic scattering rate $\sim \alpha^2T$ is larger than elastic scattering rate, and dominates the transport in the collision dominated regime $\omega \ll \alpha^2T $. A quantum Boltzmann equation leads to the conductivity
\begin{equation}
\sigma(\omega,T)=\frac{30.46T}{\alpha \log(1/\alpha)}\left[1-\left(\frac{i\omega}{T}\right)\frac{26.67}{\alpha^2 \log(1/\alpha)}\right]^{-1}
\end{equation}
The disorder induced initial growth of $\alpha$ will lead to non-monotonic temperature dependence of the inelastic scattering rate and the conductivity.

Our results are obtained for inversion symmetric systems, and do not apply for $\textrm{Hg}_{1-x}\textrm{Cd}_{x}\textrm{Te}$ due to broken inversion symmetry and the presence of additional gapless quadratic band at the $\Gamma$ point. In the presence of inversion symmetry breaking Dirac mass such as $\gamma_0\gamma_5$ and $\gamma_0\gamma_j$, even for a clean system one can find a metallic phase, if the inversion symmetry breaking mass exceeds $m$. In the supplementary material we have considered the effects of inversion symmetry breaking disorder, and the generic phase diagrams shown in Fig.2 remain qualitatively unchanged \cite{Supp}. In the strong disorder limit we have not accounted for the localization corrections for low energy diffusive modes, and such corrections can play important role in determining the more accurate scaling behavior in the strong disorder limit. The localization corrections are expected to drive a further phase transition from the CDM phase to disorder controlled insulating phase. The numerical work in Ref.~\onlinecite{Guo} and Ref.~\onlinecite{Loring}  for noninteracting problem in the strong disorder limit have showed the existence of a disorder induced topological Anderson insulator phase. Our work suggests that, akin to the conventional metal-insulator transition problem \cite{Belitz}, the interaction effects become strong in the diffusive metallic phase. The effects of strong interaction on metal-insulator transition and topological Anderson insulator will be addressed in a future publication.

P. G. was supported by NSF Grant No. DMR-1006985. S. C. was supported by NSF under the Grant DMR-1004520.

\pagebreak

\onecolumngrid

\section*{Supplementary Material for EPAPS}

\section{I. Renormalization group calculation for generic time reversal symmetric disorder} In this section we provide the details of the RG calculations for Coulomb interacting Dirac fermions in the presence of generic time reversal symmetric disorder. We begin with the disorder part of the action given by
\begin{equation}
S_D=\int d^4x \bar{\psi}\bigg[V_0(\mathbf{x})\gamma_0+\\M(\mathbf{x})\mathbb{1}_{4\times4}+V_{05}(\mathbf{x})\gamma_0\gamma_5 +V_{0j}(\mathbf{x})\gamma_0\gamma_{j}\bigg]\psi
\end{equation}
The disorder bilenears $\bar{\psi}\gamma_0\psi$ and $\bar{\psi}\psi$ always correspond to random chemical potential and random mass scattering. But, the physical significance of the other four disorder bilinears depends on the crystallographic details of the material. When the point group symmetry is such that, the Dirac $\gamma$ matrices correspond to the Dirac representation, $\bar{\psi}\gamma_0\gamma_j\psi$ corresponds to random spin orbit scattering. For this reason we will denote the coupling constant for $V_{0j}(\mathbf{x})\bar{\psi}\gamma_0\gamma_j\psi$ as $\Delta_{SO}$. In the following we will demonstrate that all the bilinears are mutually coupled, and terms like $\bar{\psi}\gamma_0\gamma_j\psi$ will be generated due to the interplay of random chemical potential and random mass disorder even if its bare strength was zero. The symmetry properties of the disorder potentials are summarized in the Table~\ref{table1}. We assume  Gaussian white noise distributions for all forms of disorder, and the appropriate distributions functions are also summarized in the Table~\ref{table1}.

\begin{figure}[htbp]
\centering
\includegraphics[scale=0.85]{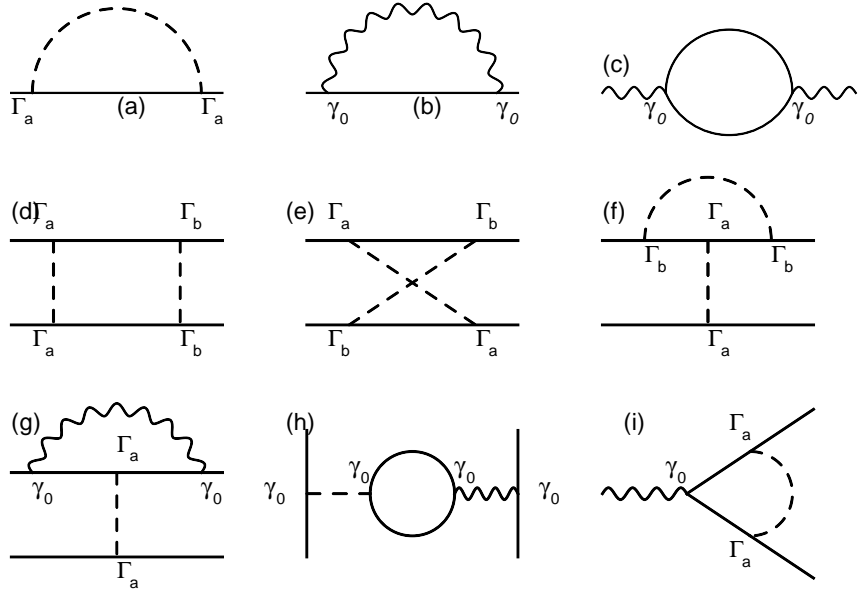}
\caption{(Color online) Relevant Feynman diagrams that do not vanish in the replica limit $n \to 0$. The disorder vertices are denoted by $\Gamma_a$ and correspond to the matrices $\gamma_0$, $\gamma_0\gamma_5$, $\mathbb{1}_{4\times4}$, and $\gamma_0\gamma_j$.}
\label{fig:supp1}
\end{figure}

\begin{table}[htdp]
\caption{The disorder distributions and their symmetries. The double angular brackets denote disorder average with respect to Gaussian white noise with zero mean.}
\begin{center}
\begin{tabular}{|l|l|l|l|l|l|}
\hline
Bilinear & $\mathcal{P}$ & $\mathcal{T}$ & $\mathcal{C}$ & $\mathcal{U}_{ch}$ & Disorder \\
\hline
\hline
$\bar{\psi}\gamma_0\psi$& + & + & - & \checkmark &$\langle\langle{V_{0}(\mathbf{x})V_{0}(\mathbf{x}^{'})}\rangle\rangle=\Delta_V\delta^{3}(\mathbf{x}-\mathbf{x}^{'})$ \\ \hline
$\bar{\psi}\gamma_0\gamma_5\psi$ & - & + & + & \checkmark & $\langle\langle{V_{05}(\mathbf{x})V_{05}(\mathbf{x}^{'})}\rangle\rangle=\Delta_{05}\delta^{3}(\mathbf{x}-\mathbf{x}^{'})$\\ \hline
$\bar{\psi}\psi$ & + & + & + & $\times$ &$\langle\langle{M(\mathbf{x})M(\mathbf{x}^{'})}\rangle\rangle=\Delta_M\delta^{3}(\mathbf{x}-\mathbf{x}^{'})$ \\ \hline
$\bar{\psi}\gamma_0\boldsymbol \gamma \psi$ & - & + & - & $\times$ & $\langle\langle{V_{0i}(\mathbf{x})V_{0j}(\mathbf{x}^{'})}\rangle\rangle=\Delta_{SO}\delta_{ij}\delta^{3}(\mathbf{x}-\mathbf{x}^{'})$\\ \hline\hline
\end{tabular}
\end{center}
\label{table1}
\end{table}
After disorder average using replica method, we obtain the following Euclidean action $\overline{S}$,
\begin{eqnarray}
&&\overline{S}=\int d^4x \left[\bar{\psi}_{a}\left\{\gamma_0(\partial_0+ig\varphi_{a})+v\gamma_{j}\partial_{j}+m -B\partial_{j}^{2}\right\}\psi_{a}+\frac{1}{2}(\partial_{j} \varphi_{a})^{2}\right]
\nonumber \\ &&-\frac{\Delta_{V}}{2}\int d^3x dx_0 dx_0^{'} \left(\bar{\psi}_{a}\gamma_{0}\psi_{a}\right)_{(x,x_0)} \left(\bar{\psi}_{b}\gamma_{0}\psi_{b}\right)_{(x,x_{0}^{'})}-\frac{\Delta_{05}}{2}\int d^3x dx_0 dx_0^{'} \left(\bar{\psi}_{a}\gamma_{0}\gamma_5\psi_{a}\right)_{(x,x_0)}\left(\bar{\psi}_{b}\gamma_{0}\gamma_5\psi_{b}\right)_{(x,x_{0}^{'})}\nonumber \\&&-\frac{\Delta_{M}}{2}\int d^3x dx_0 dx_0^{'} \left(\bar{\psi}_{a}\psi_{a}\right)_{(x,x_0)} \left(\bar{\psi}_{b}\psi_{b}\right)_{(x,x_{0}^{'})}-\frac{\Delta_{SO}}{2}\int d^3x dx_0 dx_0^{'} \left(\bar{\psi}_{a}\gamma_{0}\gamma_j\psi_{a}\right)_{(x,x_0)} \left(\bar{\psi}_{b}\gamma_{0}\gamma_j\psi_{b}\right)_{(x,x_{0}^{'})}
\label{eq:supp1}
\end{eqnarray}
Now we perform the one loop RG calculation in the momentum shell scheme, where we eliminate the fast degrees of freedom belonging to the shell $\Lambda e^{-l}<|\mathbf{k}|<\Lambda$, $-\infty<k_0<\infty$. All the relevant Feynman diagrams which do not vanish in the replica limit $n \to 0$ are shown in Fig.~\ref{fig:supp1}.

The clean, noninteracting fermion propagator $G_{0}(k_0,\mathbf{k})$ and the bare scalar potential propagator $D_{0}(k_0,\mathbf{k})$ are given by
\begin{eqnarray}
G_{0}(k_0,\mathbf{k})&=&\frac{-i(\gamma_{0}k_0+v\gamma_jk_j)+(m+Bk^2)}{k_{0}^2+v^2\mathbf{k}^2+(m+Bk^2)^2} \\
D_{0}(k_0,\mathbf{k})&=&\frac{1}{\mathbf{k}^2}
\end{eqnarray}

\paragraph{\bf{Disorder induced fermion self-energy} :}We first consider the disorder induced fermion self-energy diagrams of Fig.~\ref{fig:supp1}(a). Due to the rotational invariance, the disorder induced self-energy diagrams are independent of external momentum and depend only on the external frequency. The expression for the disorder induced self-energy is given by
\begin{eqnarray}
&&\Sigma^{D}(k_0,\mathbf{k})=\sum_{a}\Delta_{a}\int^{'}\frac{d^{3}q}{(2\pi)^3}\Gamma_{a}G_{0}(k_0,\mathbf{q})\Gamma_{a}\\
&&=\frac{(\Delta_V+\Delta_{M}+\Delta_{05}+3\Delta_{SO})\Lambda l}{(2\pi^2v^2)(1+(m+B\Lambda^2)^2v^{-2}\Lambda^{-2})}\left(-ik_0\gamma_0\right)+\frac{(\Delta_V+\Delta_{M}-\Delta_{05}-3\Delta_{SO})\Lambda l}{(2\pi^2v^2)(1+(m+B\Lambda^2)^2v^{-2}\Lambda^{-2})}(m+B\Lambda^2),
\end{eqnarray}
where $\int^{'}d^3q/(2\pi)^3$ denotes the momentum shell integration with $\Lambda e^{-l}<q<\Lambda$. The two parts of $\Sigma^{D}(k_0,\mathbf{k})$ respectively cause field and mass renormalizations due to disorder.

\paragraph{\bf{Coulomb interaction induced fermion self-energy} :} Now consider the Coulomb interaction induced exchange self-energy graph shown in Fig.~\ref{fig:supp1}(b). Due to the instantaneous approximation for the Coulomb interaction this diagram is independent of the external frequency. The expression for the exchange self-energy is given by
\begin{eqnarray}
\Sigma^{ex}(k_0,\mathbf{k})=-g^2\int_{-\infty}^{\infty} \frac{dq_0}{(2\pi)} \int^{'}\frac{d^{3}q}{(2\pi)^3}\gamma_0G_0(q_0,\mathbf{q})\gamma_0D_0(k_0-q_0,\mathbf{k}-\mathbf{q})\\
=-g^2\int_{-\infty}^{\infty} \frac{dq_0}{2\pi} \int^{'}\frac{d^{3}q}{(2\pi)^3}\frac{-i\gamma_0q_0+iv\gamma_jq_j+m+Bq^2}{\left[q_{0}^2+v^2\mathbf{q}^2+(m+Bq^2)\right](\mathbf{k}-\mathbf{q})^2}
\end{eqnarray}
Up to the quadratic order in $k$, we obtain
\begin{eqnarray}
\Sigma^{ex}(k_0,\mathbf{k})\approx -\frac{\alpha l}{2\pi \sqrt{1+(m+B\Lambda^2)^2v^{-2}\Lambda^{-2}}}\left[i\frac{4}{3}v\gamma_jk_j+2(m+B\Lambda^2)+\frac{2}{3}(m+B\Lambda^2)\Lambda^{-2}k^2\right]
\end{eqnarray}

The disorder and instantaneous Coulomb interaction destroy the Lorentz invariance of the noninteracting action, and this is manifested in the inequality of the coefficients of $\gamma_0 k_0$ and $\gamma_j k_j$. Together these terms give rise to a multiplicative field renormalization constant $Z_{\psi}$, and a scale dependent dynamic scaling exponent $z(l)$. Since the field renormalization at one loop order arises only from the disorder contribution, Ward identity ensures that the renormalization of the Coulomb interaction vertex comes solely from the disorder contribution, shown in Fig.~\ref{fig:supp1}(i).

\paragraph{\bf {Scalar potential self-energy}:} We next consider the scalar potential self-energy diagram shown in Fig.~\ref{fig:supp1}(c).
The scalar potential self-energy at one loop order is given by
\begin{eqnarray}
&&\Pi(k_0,\mathbf{k})=g^2\int_{-\infty}^{\infty} \frac{dq_0}{(2\pi)} \int^{'}\frac{d^{3}q}{(2\pi)^3}Tr[\gamma_0G_0(q_0,\mathbf{q})\gamma_0G_0(k_0+q_0,\mathbf{k}+\mathbf{q})] \\&&=-4g^2\int_{-\infty}^{\infty} \frac{dq_0}{(2\pi)} \int^{'}\frac{d^{3}q}{(2\pi)^3}\frac{q_0(k_0+q_0)-v^2\mathbf{q}\cdot(\mathbf{k}+\mathbf{q})-(m+Bq^2)(m+B(\mathbf{k}+\mathbf{q})^2)}{\left[q_{0}^2+v^2\mathbf{q}^2+(m+B\mathbf{q}^2)^2\right]\left[(k_{0}+q_{0})^2+v^2(\mathbf{k}+\mathbf{q})^2+(m+B(\mathbf{k}+\mathbf{q})^2)^2\right]}\nonumber \\
\end{eqnarray}
After some simple integrations we can show $\Pi(0,0)=0$ and to the quadratic order in momentum and frequency we obtain
\begin{equation}
\Pi(k_0,\mathbf{k})\approx-\frac{2\alpha l\mathbf{k}^2}{3\pi}\frac{1+\frac{3}{2}(m^2+B^2\Lambda^4)v^{-2}\Lambda^{-2}+mBv^{-2}}{\left[1+(m+B\Lambda^2)^2v^{-2}\Lambda^{-2}\right]^{\frac{5}{2}}}
\end{equation}
The scalar potential self-energy causes a field renormalization of the scalar field $\varphi$, and which in turn leads to the charge renormalization.

\paragraph{\bf {Renormalization constants for} $\bf v$, $\bf m$, $\bf B$ and $\boldsymbol \alpha$ :}
At this point we introduce dimensionless parameters $m\to mv^{-1}\Lambda^{-1}$, $B \to B\Lambda v^{-1}$, and $\Delta_a \to (\Delta_a \Lambda)/(2\pi^2v^2)$. After collecting all the self-energies, and the Coulomb vertex corrections, and performing the anisotropic rescaling of the space-time coordinates as, $x_0 \to x_0 e^{zl}$, $\mathbf{x}\to \mathbf{x} e^{l}$, the parts of action without the disorder vertex become
\begin{eqnarray}
\bar{S}^{'}&=&\int d^4x e^{(z+3)l}\bigg[\bar{\psi}_a\bigg\{\left(1+\frac{(\Delta_V+\Delta_M+\Delta_{05}+3\Delta_{SO})l}{1+(m+B)^2}\right)\gamma_0(e^{-zl}\partial_0+ig\varphi_{a})
+ve^{-l}\left(1+\frac{2\alpha l}{3\pi \sqrt{1+(m+B)^2}}\right)\gamma_j\partial_j\nonumber \\&&+v\Lambda\left(m-\frac{(\Delta_V+\Delta_M-\Delta_{05}-3\Delta_{SO}))l}{1+(m+B)^2}(m+B)+\frac{\alpha l}{\pi \sqrt{1+(m+B)^2}}(m+B)\right)\nonumber \\&&-v\Lambda^{-1}\bigg(B+\frac{\alpha l}{3\pi \sqrt{1+(m+B)^2}}(m+B)\bigg)\times e^{-2l}\partial_{j}^{2}
\bigg\}\psi_a+\frac{1}{2}e^{-2l}\left(1+\frac{2\alpha l}{3\pi}\frac{1+\frac{3}{2}(m^2+B^2)+mB}{\left[1+(m+B)^2\right]^{\frac{5}{2}}}\right)(\partial_j \varphi_{a})^{2}\bigg]\nonumber \\
\end{eqnarray}
Now we introduce the renormalization constants for the fermionic field $\psi$, the scalar field $\varphi$, the charge $g$, the Fermi velocity $v$, the mass parameter $m$ and the higher derivative parameter $B$ according to $\psi \to Z_{\psi}^{-1/2}\psi$, $\varphi \to Z_{\varphi}^{-1/2} \varphi$, $g \to Z_{g}^{-1/2}g$, $v \to Z_{v}^{-1} v$, $\alpha \to Z_{g}^{-1}Z_v \alpha=Z^{-1}_{\alpha}\alpha$, $m \to Z_{m}^{-1}m$ and $B \to Z_{B}^{-1} B$, to recast the action in the original form, and find
\begin{eqnarray}
Z_{\psi}&=&e^{3l}\left[1+\frac{(\Delta_V+\Delta_M+\Delta_{05}+3\Delta_{SO})l}{1+(m+B)^2}\right]\\
Z_{\varphi}&=&e^{(z+1)l}\left[1+\frac{2\alpha l}{3\pi}\frac{1+\frac{3}{2}(m^2+B^2)+mB}{\left[1+(m+B)^2\right]^{\frac{5}{2}}}\right]\\
Z_{g}&=&e^{(z-1)l}\left[1-\frac{2\alpha l}{3\pi}\frac{1+\frac{3}{2}(m^2+B^2)+mB}{\left[1+(m+B)^2\right]^{\frac{5}{2}}}\right]\\
Z_v&=&e^{(z-1)l}\left[1+\frac{2\alpha l}{3\pi \sqrt{1+(m+B)^2}}-\frac{(\Delta_V+\Delta_M+\Delta_{05}+3\Delta_{SO})l}{1+(m+B)^2}\right]\\
Z_m&=&e^l\left[1-\frac{2\alpha l}{3\pi \sqrt{1+(m+B)^2}}+\left \{\frac{\alpha l}{\pi \sqrt{1+(m+B)^2}}-\frac{(\Delta_V+\Delta_M-\Delta_{05}-3\Delta_{SO})l}{1+(m+B)^2}\right \}\left(1+\frac{B}{m}\right)\right]\\
Z_B&=&e^{-l}\left[1-\frac{2\alpha l}{3\pi \sqrt{1+(m+B)^2}}+\frac{\alpha l}{\pi \sqrt{1+(m+B)^2}}\left(1+\frac{B}{m}\right)\right]
\end{eqnarray}
Using the renormalization constants given above, we obtain the RG flow equations for $v$, $m$, $B$ and $\alpha$. Next we consider the renormalization of the disorder vertices due to the interplay of disorder and Coulomb interaction. The relevant diagrams are shown in Fig.~\ref{fig:supp1}(d), Fig.~\ref{fig:supp1}(e), Fig.~\ref{fig:supp1}(f), Fig.~\ref{fig:supp1}(g), Fig.~\ref{fig:supp1}(h). For the renormalization of the disorder vertices we can set all the external frequencies and momenta to zero in these Feynman diagrams.

\paragraph{\bf{Renormalization of} $\boldsymbol \Delta_{\bf V}$ :} The renormalization of $\Delta_V$ from graphs Fig.~\ref{fig:supp1}(d), Fig.~\ref{fig:supp1}(e), arise due to the interplay of the pairs $(\mathbb{1}_{4\times4},\gamma_0\gamma_j)$ and $(\gamma_0,\mathbb{1}_{4\times4} )$. After accounting for the symmetry factor of two for each graph, we find the total contribution
\begin{equation}
\delta \Delta_{V}^{1(d)+1(e)}=\frac{4\Delta_{M}\Delta_{SO}\Lambda l}{(2\pi^2v^2)(1+(m+B\Lambda^2)^2v^{-2}\Lambda^{-2})^2}+\frac{4\Delta_V\Delta_{M}\Lambda l}{(2\pi^2v^2)(1+(m+B\Lambda^2)^2v^{-2}\Lambda^{-2})^2}\left(\frac{(m+B\Lambda^2)^2}{v^2\Lambda^2}\right)
\end{equation}
For the diagrams corresponding to Fig.~\ref{fig:supp1}(f), we take $\Gamma_a=\gamma_0$, and $\Gamma_b=\gamma_0, \ \mathbb{1}_{4\times4}, \ \gamma_0\gamma_5, \ \gamma_0\gamma_i$, and after accounting for a symmetry factor of two, we find
\begin{equation}
\delta \Delta_{V}^{1(f)}=2\left[\Delta_{V}^{2}+\Delta_V\Delta_M+\Delta_V\Delta_{05}+3\Delta_V\Delta_{SO}\right]\frac{\Lambda l}{(2\pi^2v^2)(1+(m+B\Lambda^2)^2v^{-2}\Lambda^{-2})}
\end{equation}
The graphs in Fig.~\ref{fig:supp1}(g), and Fig.~\ref{fig:supp1}(h) represent the renormalization of disorder couplings due to Coulomb interaction. Notice that Fig.~\ref{fig:supp1}(h) that represents a screening effect due to the presence of internal fermion loop, occurs only for random chemical potential. In Fig.~\ref{fig:supp1}(g) we set $\Gamma_a=\gamma_0$, and find
\begin{equation}
\delta \Delta_{V}^{1(g)}=2g^2\Delta_{V}\int_{-\infty}^{\infty} \frac{dq_0}{(2\pi)} \int^{'}\frac{d^{3}q}{(2\pi)^3}\frac{q_{0}^{2}-v^2\mathbf{q}^2-(m+B\mathbf{q}^2)^2}{(q_{0}^{2}+v^2\mathbf{q}^{2}+(mv\Lambda+Bv\Lambda^{-1}\mathbf{q}^2)^2)^2\mathbf{q}^2}=0
\end{equation}
After accounting for the symmetry factor of two, Fig.~\ref{fig:supp1}(h) leads to
\begin{equation}
\delta \Delta_{V}^{1(h)}=-\frac{4\alpha l}{3\pi}\frac{1+\frac{3}{2}(m^2+B^2\Lambda^4)v^{-2}\Lambda^{-2}+mBv^{-2}}{\left[1+(m+B\Lambda^2)^2v^{-2}\Lambda^{-2}\right]^{\frac{5}{2}}}
\end{equation}
After collecting all the perturbative corrections to $\Delta_V$, we switch to dimensionless couplings, and introduce the renormalization constant $Z_{\Delta_V}$. We find
\begin{eqnarray}
Z_{\Delta_V}&=&e^{-l}\bigg[1+\frac{2(\Delta_V+\Delta_{05}+3\Delta_{SO})l}{1+(m+B)^2}+\frac{2\Delta_Ml}{\left(1+(m+B)^2\right)^2}\left\{1+3(m+B)^2+\frac{2\Delta_{SO}}{\Delta_V}\right\}-\frac{4\alpha l}{3\pi \sqrt{1+(m+B)^2}}\nonumber \\ &&-\frac{4\alpha l}{3\pi}\frac{1+\frac{3}{2}(m^2+B^2)+mB}{\left[1+(m+B)^2\right]^{\frac{5}{2}}}\bigg]
\end{eqnarray}

\paragraph{\bf{Renormalization of} $\boldsymbol \Delta_{\bf{M}}$ :} The renormalization of $\Delta_M$ from graphs Fig.~\ref{fig:supp1}(d), Fig.~\ref{fig:supp1}(e), arise due to the interplay of pairs ($\gamma_0$, $\gamma_0$), ($\mathbb{1}_{4\times4}$,$\mathbb{1}_{4\times4}$), ($\gamma_0\gamma_5$,$\gamma_0\gamma_5$), ($\gamma_0\gamma_j$,$\gamma_0\gamma_j$), and ($\gamma_0$,$\gamma_0\gamma_j$). After accounting for the symmetry factor of two for each graph, we find the total contribution
\begin{equation}
\delta \Delta_{M}^{1(d)+1(e)}=\frac{4(\Delta_{V}^{2}+\Delta_{M}^{2}+\Delta_{05}^{2}+3\Delta_{SO}^{2})\Lambda l}{(2\pi^2v^2)(1+(m+B\Lambda^2)^2v^{-2}\Lambda^{-2})^2}\left(\frac{(m+B\Lambda^2)^2}{v^2\Lambda^2}\right)+\frac{4\Delta_{V}\Delta_{SO}\Lambda l}{(2\pi^2v^2)(1+(m+B\Lambda^2)^2v^{-2}\Lambda^{-2})^2}
\end{equation}
For the diagrams corresponding to Fig.~\ref{fig:supp1}(f), we take $\Gamma_a=\mathbb{1}_{4\times4}$, and $\Gamma_b=\gamma_0, \ \mathbb{1}_{4\times4}, \ \gamma_0\gamma_5, \ \gamma_0\gamma_i$, and after accounting for a symmetry factor of two, we find
\begin{equation}
\delta \Delta_{M}^{1(f)}=2(-\Delta_{M}^{2}-\Delta_M\Delta_V+\Delta_M\Delta_{05}+3\Delta_M\Delta_{SO})\frac{\Lambda l}{(2\pi^2v^2)(1+(m+B\Lambda^2)^2v^{-2}\Lambda^{-2})^2}\left(1-\frac{(m+B\Lambda^2)^2}{v^2\Lambda^2}\right)
\end{equation}
In Fig.~\ref{fig:supp1}(g) we set $\Gamma_a=\mathbb{1}_{4\times4}$, and accounting for the symmetry factor of two we find
\begin{equation}
\delta \Delta_{M}^{1(g)}=\frac{2\alpha \Delta_Ml}{\pi \left(1+(m+B)^2\right)^{\frac{3}{2}}}
\end{equation}
After collecting all the perturbative corrections to $\Delta_{M}$, we switch to dimensionless couplings, and introduce the renormalization constant $Z_{\Delta_{M}}$. We find
\begin{eqnarray}
&&Z_{\Delta_M}=e^{-l}\bigg[1-2\Delta_Ml\frac{1-3(m+B)^2}{\left(1+(m+B)^2\right)^2}-2(\Delta_V-\Delta_{05}-3\Delta_{SO})l\frac{1-(m+B)^2}{\left(1+(m+B)^2\right)^2}-\frac{4\alpha l }{3\pi \sqrt{1+(m+B)^2}}\nonumber \\&& +\frac{2\alpha l}{\pi \left(1+(m+B)^2\right)^{\frac{3}{2}}} +\frac{4\Delta_V\Delta_{SO}l}{\left(1+(m+B)^2\right)^2}+4\left(\Delta_{V}^{2}+\Delta_{05}^{2}+3\Delta_{SO}^{2}\right)l\frac{(m+B)^2}{\left(1+(m+B)^2\right)^2}\bigg]
\end{eqnarray}

\paragraph{\bf{Renormalization of} $\boldsymbol \Delta_{\bf{05}}$ :} The renormalization of $\Delta_{05}$ from graphs Fig.~\ref{fig:supp1}(d), Fig.~\ref{fig:supp1}(e), arise due to the interplay of pairs ($\gamma_0\gamma_i$, $\gamma_0\gamma_j$ with $i\neq j$) and ($\mathbb{1}_{4\times4}$,$\gamma_0\gamma_5$). After accounting for the symmetry factor of two for each graph, we find the total contribution
\begin{equation}
\delta \Delta_{05}^{1(d)+1(e)}=\frac{4\Delta_{SO}^{2}\Lambda l}{(2\pi^2v^2)(1+(m+B\Lambda^2)^2v^{-2}\Lambda^{-2})^2}+\frac{4\Delta_{05}\Delta_{M}\Lambda l}{(2\pi^2v^2)(1+(m+B\Lambda^2)^2v^{-2}\Lambda^{-2})^2}\left(\frac{(m+B\Lambda^2)^2}{v^2\Lambda^2}\right)
\end{equation}
For the diagrams corresponding to Fig.~\ref{fig:supp1}(f), we take $\Gamma_a=\gamma_0\gamma_5$, and $\Gamma_b=\gamma_0, \ \mathbb{1}_{4\times4}, \ \gamma_0\gamma_5, \ \gamma_0\gamma_i$, and after accounting for a symmetry factor of two, we find
\begin{eqnarray}
\delta \Delta_{05}^{1(f)}=\frac{2\left[\Delta_{05}^{2}+\Delta_{05}\Delta_V-\Delta_{05}\Delta_{M}\right]\Lambda l}{(2\pi^2v^2)(1+(m+B\Lambda^2)^2v^{-2}\Lambda^{-2})^2}\left(1-\frac{(m+B\Lambda^2)^2}{v^2\Lambda^2}\right)-\frac{6\Delta_{05}\Delta_{SO}\Lambda l}{(2\pi^2v^2)(1+(m+B\Lambda^2)^2v^{-2}\Lambda^{-2})^2}
\end{eqnarray}
In Fig.~\ref{fig:supp1}(g) we set $\Gamma_a=\gamma_0\gamma_5$, and accounting for the symmetry factor of two we find
\begin{equation}
\delta \Delta_{05}^{1(g)}=\frac{2\alpha l}{\pi \left(1+(m+B)^2\right)^{\frac{3}{2}}}\left(\frac{(m+B\Lambda^2)^2}{v^2\Lambda^2}\right)
\end{equation}
After collecting all the perturbative corrections to $\Delta_{05}$, we switch to dimensionless couplings, and introduce the renormalization constant $Z_{\Delta_{05}}$. We find
\begin{eqnarray}
Z_{\Delta_{05}}&=&e^{-l}\bigg[1+2(\Delta_{05}+\Delta_V)l\frac{1-(m+B)^2}{\left(1+(m+B)^2\right)^2}-2\Delta_Ml\frac{1-3(m+B)^2}{\left(1+(m+B)^2\right)^2}-\frac{6\Delta_{SO}l}{\left(1+(m+B)^2\right)^2}\nonumber \\&&-\frac{4\alpha l }{3\pi \sqrt{1+(m+B)^2}}+\frac{2\alpha l(m+B)^2}{\pi \left(1+(m+B)^2\right)^{\frac{3}{2}}} +\frac{4\Delta_{SO}^{2}l}{\Delta_{05}\left(1+(m+B)^2\right)^2}\bigg]
\end{eqnarray}

\paragraph{\bf{Renormalization of} $\boldsymbol \Delta_{\mathbf{SO}}$ :} The renormalization of $\Delta_{SO}$ from graphs Fig.~\ref{fig:supp1}(d), Fig.~\ref{fig:supp1}(e), arise due to the interplay of pairs ($\gamma_0\gamma_5$,$\gamma_0\gamma_i$), ($\gamma_0$, $\mathbb{1}_{4\times4}$), ($\gamma_0\gamma_i$, $\mathbb{1}_{4\times4}$). These diagrams possess linear UV divergence, and after accounting for the symmetry factor of two for each graph, we find the total contribution
\begin{equation}
\delta \Delta_{SO}^{1(d)+1(e)}=\frac{4(2\Delta_{SO}\Delta_{05}+\Delta_V\Delta_M)\Lambda l}{3(2\pi^2v^2)(1+(m+B\Lambda^2)^2v^{-2}\Lambda^{-2})^2}+\frac{4\Delta_{SO}\Delta_{M}\Lambda l}{(2\pi^2v^2)(1+(m+B\Lambda^2)^2v^{-2}\Lambda^{-2})^2}\left(\frac{(m+B\Lambda^2)^2}{v^2\Lambda^2}\right)
\end{equation}
For the diagrams corresponding to Fig.~\ref{fig:supp1}(f), we take $\Gamma_a=\gamma_0\gamma_i$, and $\Gamma_b=\gamma_0, \ \mathbb{1}_{4\times4}, \ \gamma_0\gamma_5, \ \gamma_0\gamma_j$, and after accounting for a symmetry factor of two, we find
\begin{equation}
\delta \Delta_{SO}^{1(f)}=-\frac{2\left[\Delta_{SO}^{2}-\Delta_{SO}\Delta_V+\Delta_{SO}\Delta_{M}\right]\Lambda l}{3(2\pi^2v^2)(1+(m+B\Lambda^2)^2v^{-2}\Lambda^{-2})^2}\left(1-\frac{3(m+B\Lambda^2)^2}{v^2\Lambda^2}\right)-\frac{2\Delta_{05}\Delta_{SO}\Lambda l}{3(2\pi^2v^2)(1+(m+B\Lambda^2)^2v^{-2}\Lambda^{-2})^2}
\end{equation}
In Fig.~\ref{fig:supp1}(g) we set $\Gamma_a=\gamma_0\gamma_i$, and accounting for the symmetry factor of two we find
\begin{equation}
\delta \Delta_{SO}^{1(g)}=\frac{2\alpha \Delta_{SO} l}{3\pi \left(1+(m+B)^2\right)^{\frac{3}{2}}}\left(1+3\frac{(m+B\Lambda^2)^2}{v^2\Lambda^2}\right)
\end{equation}
After collecting all the perturbative corrections to $\Delta_{SO}$, we switch to dimensionless couplings, and introduce the renormalization constant $Z_{\Delta_{SO}}$. We find
\begin{eqnarray}
Z_{\Delta_{SO}}&=&e^{-l}\bigg[1-\frac{2l}{3}(\Delta_{SO}-\Delta_V)\frac{1-3(m+B)^2}{\left(1+(m+B)^2\right)^2}-\frac{2\Delta_Ml}{3}\frac{1-9(m+B)^2}{\left(1+(m+B)^2\right)^2}+\frac{2\Delta_{05}l}{1+(m+B)^2}\nonumber \\&& +\frac{2\alpha(1+3(m+B)^2) l}{3\pi \left(1+(m+B)^2\right)^{\frac{3}{2}}}-\frac{4\alpha l}{3\pi \sqrt{1+(m+B)^2}}+\frac{4\Delta_V\Delta_M l}{3\Delta_{SO}\left(1+(m+B)^2\right)^2}\bigg]
\end{eqnarray}

\paragraph{\bf{RG flow equations}:}Using the renormalization constants found for the dimensionless coupling constants we obtain the following RG flow equations
\begin{eqnarray}
&&\frac{dv}{dl}=v\left[z-1+\frac{2\alpha }{3\pi \sqrt{1+(m+B)^2}}-\frac{(\Delta_V+\Delta_M+\Delta_{05}+3\Delta_{SO})}{1+(m+B)^2}\right]\\
&&\frac{dm}{dl}=m+(m+3B)\frac{\alpha }{3\pi \sqrt{1+(m+B)^2}}-(m+B)\frac{(\Delta_V+\Delta_M-\Delta_{05}-3\Delta_{SO})}{1+(m+B)^2}\\
&&\frac{dB}{dl}=-B+(m-B)\frac{\alpha }{3\pi \sqrt{1+(m+B)^2}}\\
&&\frac{d\alpha}{dl}=\alpha\left[\frac{(\Delta_V+\Delta_M+\Delta_{05}+3\Delta_{SO})}{1+(m+B)^2}-\frac{2\alpha }{3\pi \sqrt{1+(m+B)^2}}-\frac{2\alpha }{3\pi}\frac{1+\frac{3}{2}(m^2+B^2)+mB}{\left[1+(m+B)^2\right]^{\frac{5}{2}}}\right]\\
&&\frac{d\Delta_V}{dl}=\Delta_V\bigg[-1+\frac{2(\Delta_V+\Delta_{05}+3\Delta_{SO})}{1+(m+B)^2}+2\Delta_M\frac{1+3(m+B)^2}{\left(1+(m+B)^2\right)^2}-\frac{4\alpha }{3\pi \sqrt{1+(m+B)^2}}\nonumber \\ &&-\frac{4\alpha }{3\pi}\frac{1+\frac{3}{2}(m^2+B^2)+mB}{\left[1+(m+B)^2\right]^{\frac{5}{2}}}\bigg]+\frac{4\Delta_M\Delta_{SO}}{\left(1+(m+B)^2\right)^2}\\
&&\frac{d\Delta_M}{dl}=\Delta_M\bigg[-1-2\Delta_M\frac{1-3(m+B)^2}{\left(1+(m+B)^2\right)^2}-2(\Delta_V-\Delta_{05}-3\Delta_{SO})\frac{1-(m+B)^2}{\left(1+(m+B)^2\right)^2}-\frac{4\alpha }{3\pi \sqrt{1+(m+B)^2}}\nonumber \\&& +\frac{2\alpha }{\pi \left(1+(m+B)^2\right)^{\frac{3}{2}}} \bigg]+\frac{4\Delta_V\Delta_{SO}}{\left(1+(m+B)^2\right)^2}+4\left(\Delta_{V}^{2}+\Delta_{05}^{2}+3\Delta_{SO}^{2}\right)\frac{(m+B)^2}{\left(1+(m+B)^2\right)^2}\\
&&\frac{d\Delta_{05}}{dl}=\Delta_{05}\bigg[-1+2(\Delta_{05}+\Delta_V)\frac{1-(m+B)^2}{\left(1+(m+B)^2\right)^2}-2\Delta_M\frac{1-3(m+B)^2}{\left(1+(m+B)^2\right)^2}-\frac{6\Delta_{SO}}{\left(1+(m+B)^2\right)^2}\nonumber \\&&-\frac{4\alpha }{3\pi \sqrt{1+(m+B)^2}}+\frac{2\alpha(m+B)^2}{\pi \left(1+(m+B)^2\right)^{\frac{3}{2}}} \bigg]+\frac{4\Delta_{SO}^{2}}{\left(1+(m+B)^2\right)^2}\\
&&\frac{d\Delta_{SO}}{dl}=\Delta_{SO}\bigg[-1-\frac{2}{3}(\Delta_{SO}-\Delta_V)\frac{1-3(m+B)^2}{\left(1+(m+B)^2\right)^2}-\frac{2\Delta_M}{3}\frac{1-9(m+B)^2}{\left(1+(m+B)^2\right)^2}+\frac{2\Delta_{05}}{1+(m+B)^2}\nonumber \\&& +\frac{2\alpha (1+3(m+B)^2)}{\pi \left(1+(m+B)^2\right)^{\frac{3}{2}}}-\frac{4\alpha }{3\pi \sqrt{1+(m+B)^2}}\bigg]+\frac{4\Delta_V\Delta_M}{3\left(1+(m+B)^2\right)^2}
\end{eqnarray}
The RG equations presented in the main text are obtained by setting $\Delta_M=\Delta_{05}=\Delta_{SO}=0$. Notice that in the presence of a finite mass or $B$, a random mass is always generated from the other disorders at quadratic order.

\paragraph{\bf{Fixed point analysis} :} The RG flow equations for generic time reversal symmetric disorder have following five fixed points
\begin{eqnarray}
&&\mathrm{FP1}:\Delta_{V}^{\ast}=\Delta_{M}^{\ast}=\Delta_{05}^{\ast}=\Delta_{SO}^{\ast}=\alpha^{\ast}=m=B=0, \ z=1\\
&&\mathrm{FP2}: \Delta_{V}^{\ast}+\Delta_{05}^{\ast}=\frac{1}{2}, \ \Delta_{M}^{\ast}=\Delta_{SO}^{\ast}=\alpha^{\ast}=m=B=0, \ z=\frac{3}{2}\\
&&\mathrm{FP3}: \Delta_{05}^{\ast}-\Delta_{M}^{\ast}=\frac{1}{2}, \ \Delta_{V}^{\ast}=\Delta_{SO}^{\ast}=\alpha^{\ast}=m=B=0, \ z=\frac{3}{2}+2\Delta_{M}^{\ast}\\
&&\mathrm{FP4}: \Delta_{05}^{\ast}=\frac{9}{10}, \ \Delta_{SO}^{\ast}=\frac{6}{5}, \ \Delta_{V}^{\ast}=\Delta_{M}^{\ast}=\alpha^{\ast}=m=B=0,  z=\frac{31}{10}\\
&&\mathrm{FP5}:\Delta_{05}^{\ast}=\frac{4\alpha^{\ast}}{3\pi}=1, \ \Delta_{V}^{\ast}= \Delta_{M}^{\ast}=\Delta_{SO}^{\ast}=m=B=0, \ z=\frac{3}{2}
\end{eqnarray}
The fixed points FP2, FP3, FP4, FP5 describe possible universality classes of the SM-massless CDM phase transitions.

FP1 is the noninteracting, clean fixed point. Upto a critical strength of disorder this fixed point is stable, and the fermion mass is the only relevant perturbation. In its vicinity $\Delta_a \approx \Delta_{a0} e^{-l}$. To see how the TI-BI phase boundary is shifted by various disorders, we first consider the noninteracting problem by setting $\alpha=0$. For $B$, we have $B= B_0e^{-l}$. The flow equation for mass $m$ can be approximated as
\begin{equation}
\frac{dm}{dl}\approx m-B_0(\Delta_{V0}+\Delta_{M0}-\Delta_{05,0}-3\Delta_{SO,0})e^{-2l}
\end{equation}
which has the solution
\begin{equation}
m-\frac{B}{3}(\Delta_V+\Delta_{M}-\Delta_{05}-3\Delta_{SO})=\left[m_0-\frac{B}{3}(\Delta_{V0}+\Delta_{M0}-\Delta_{05,0}-3\Delta_{SO,0})\right]e^{l}
\end{equation}
The phase boundary shifts to $m=\frac{B}{3}(\Delta_{V}+\Delta_{M}-\Delta_{05}-3\Delta_{SO})$. If $\Delta_V+\Delta_M>\Delta_{05}+3\Delta_{SO}$, TI has a larger regime of stability. In the presence of Coulomb interaction we find a further shift of the phase boundary to $m=\frac{B}{3}(\Delta_{V}+\Delta_{M}-\Delta_{05}-3\Delta_{SO})-\frac{B\alpha}{2\pi}$. In the vicinity of this fixed point $\alpha$ is marginally irrelevant, and decreases logarithmically.

\begin{figure}[ht]
\centering
\subfigure[]{
\includegraphics[scale=0.85]{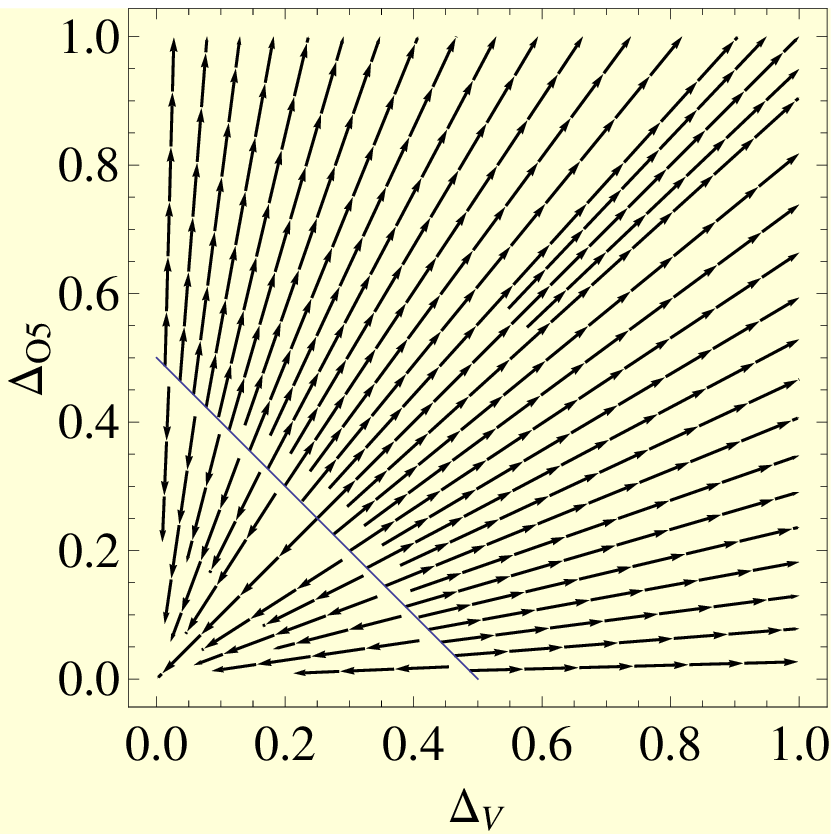}
\label{fig:subfigsupp2a}
}
\subfigure[]{
\includegraphics[scale=0.85]{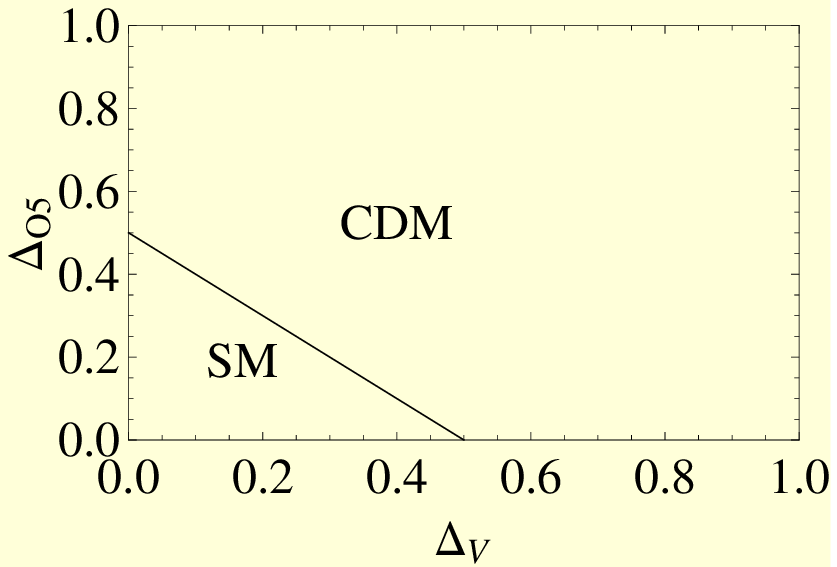}
\label{fig:subfigsupp2b}
}
\label{fig:supp2}
\caption[]{(a) The RG flow, and (b) the phase diagram in $\Delta_V-\Delta_{05}$ plane for $m=B=\alpha=\Delta_M=\Delta_{SO}=0$. The blue line $\Delta_V+\Delta_{05}=\frac{1}{2}$ corresponds to the SM-CDM phase boundary.
}
\end{figure}

\begin{figure}[t!]
\centering\includegraphics[scale=0.85]{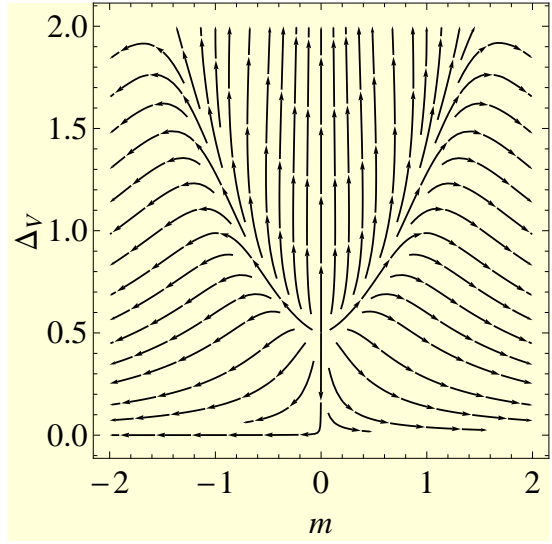}
\caption{(Color online) RG flow in $m-\Delta_V$ plane for $B=\alpha=\Delta_M=\Delta_{05}=\Delta_{SO}=0$.
 }
\label{fig:supp3}
\end{figure}

The line of fixed points FP2, describes the phase boundary between the SM and massless CDM phases for the noninteracting problem with chiral symmetric disorder. The associated RG flow and phase diagram in $\Delta_V-\Delta_{05}$ plane, for $m=B=\alpha=\Delta_M=\Delta_{SO}=0$ are respectively shown in Fig.~\ref{fig:subfigsupp2a} and Fig.~\ref{fig:subfigsupp2b}.
Now we linearize the recursion relations in the vicinity of this line of fixed points, and obtain
\begin{eqnarray}
\frac{d\delta \alpha}{dl}&=&\frac{1}{2}\delta \alpha \Rightarrow \delta \alpha =\delta \alpha_0 e^{l/2}\label{eq:FP21}\\
\frac{d\delta\Delta_{M}}{dl}&=&-4\Delta_{V}^{\ast}\delta\Delta_{M}+4\Delta_{V}^{\ast}\delta\Delta_{SO}\label{eq:FP22}\\
\frac{d\delta\Delta_{SO}}{dl}&=&-\frac{4}{3}\Delta_{V}^{\ast}\delta\Delta_{SO}+\frac{4}{3}\Delta_{V}^{\ast}\delta\Delta_{M}\label{eq:FP23}\\
\frac{d\delta\Delta_{V}}{dl}&=&2\Delta_{V}^{\ast}(\delta\Delta_{V}+\delta\Delta_{05}+\delta\Delta_{M}+3\Delta_{SO}-\frac{4\delta \alpha}{3\pi})\label{eq:FP24}\\
\frac{d\delta\Delta_{05}}{dl}&=&2\Delta_{05}^{\ast}(\delta\Delta_{V}+\delta\Delta_{05}-\delta\Delta_{M}-3\Delta_{SO}-\frac{2\delta \alpha}{3\pi})\label{eq:FP25}
\end{eqnarray}
From Eq.~\ref{eq:FP22} and Eq.~\ref{eq:FP23}, we find $\delta\Delta_{M}+3\delta\Delta_{SO}=\delta\Delta_{M}^{0}+3\delta\Delta_{SO}^{0}$. Therefore, chiral symmetry breaking perturbations in the diffusive phase will remain finite. After adding both sides of Eq.~\ref{eq:FP24} and Eq.~\ref{eq:FP25} we find
\begin{eqnarray}
\delta\Delta_{V}+\delta\Delta_{05}+2(\delta\Delta_{M}+3\delta\Delta_{SO})(\Delta_{V}^{\ast}-\Delta_{05}^{\ast})-\frac{8\delta \alpha}{3\pi}(2\Delta_{V}^{\ast}+\Delta_{05}^{\ast})
&=&[\delta\Delta_{V}^{0}+\delta\Delta_{05}^{0}+2(\delta\Delta_{M}^{0}+3\delta\Delta_{SO}^{0})(\Delta_{V}^{\ast}-\Delta_{05}^{\ast})\nonumber \\&&-\frac{8\delta \alpha^{0}}{3\pi}(2\Delta_{V}^{\ast}+\Delta_{05}^{\ast})]e^{l}
\end{eqnarray}
is the most relevant variable with eigenvalue one. Therefore the mean free path diverges with exponent $\nu=1$. In the vicinity of this fixed point, the relevant variable also describes the phase boundary in the multidimensional coupling constant space, and $z$ changes continuously along the phase boundary. First we note that interaction shifts the phase boundary to higher values of $\Delta_{V}$ and $\Delta_{05}$. For $\Delta_{V}^{\ast}>\Delta_{05}^{\ast}$ the mass and the spin orbit disorders shift the phase boundary to smaller values of $\Delta_{V}$ and $\Delta_{05}$. For $\Delta_{V}^{\ast}<\Delta_{05}^{\ast}$ the mass and the spin orbit disorders shift the phase boundary to higher values of $\Delta_{V}$ and $\Delta_{05}$. Now consider $m, B\neq0$,
\begin{eqnarray}
\frac{dB}{dl}&=&-B\\
\frac{dm}{dl}&=&m\left(\frac{3}{2}-2\Delta_{V}^{\ast}\right)-B\left(2\Delta_{V}^{\ast}-\frac{1}{2}\right)
\end{eqnarray}
Notice that apart from a redefinition of the effective mass, the scaling dimension of $m$ has changed into $(3/2-2\Delta_{V}^{\ast})$. This has important role in governing the insulator-CDM phase boundaries. For $\Delta_{05}^{\ast}>\Delta_{V}^{\ast}$, the scaling dimension of $m$ is bigger than unity, and this increases the stability of the insulating phases with respect to the CDM. The RG flow in the $m-\Delta_V$ plane for $B=\alpha=\Delta_M=\Delta_{05}=\Delta_{SO}=0$ is shown in Fig.~\ref{fig:supp3}.

\begin{figure}[ht]
\centering
\subfigure[]{
\includegraphics[scale=0.85]{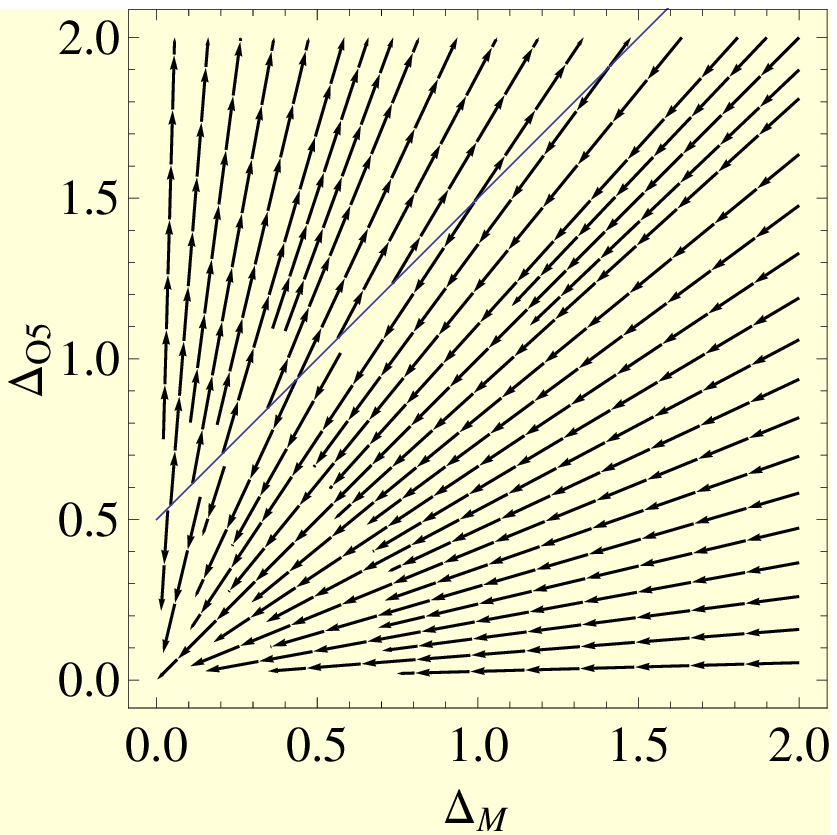}
\label{fig:subfigsupp4a}
}
\subfigure[]{
\includegraphics[scale=0.85]{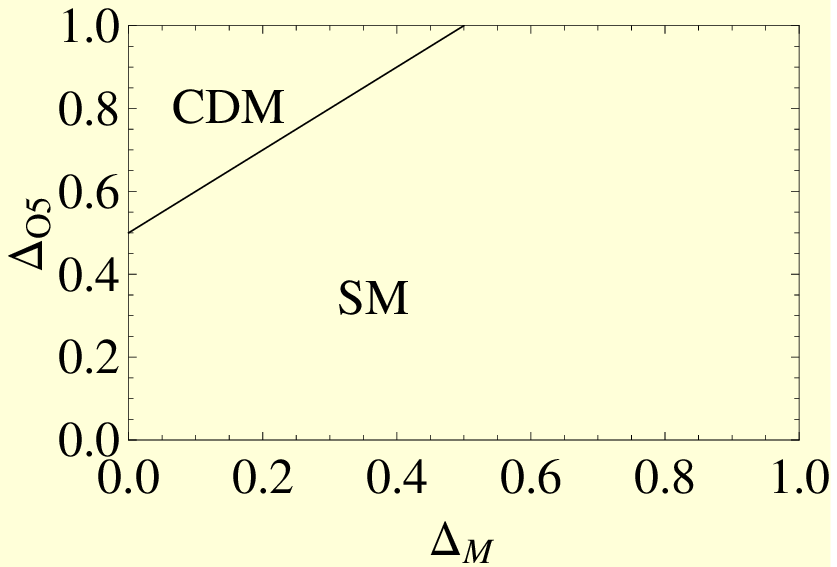}
\label{fig:subfigsupp4b}
}
\label{fig:supp4}
\caption[]{(a) The RG flow, and (b)the phase diagram in $\Delta_M-\Delta_{05}$ plane for $m=B=\alpha=\Delta_V=\Delta_{SO}=0$. The blue line $\Delta_{05}-\Delta_M=\frac{1}{2}$ corresponds to the SM-massless CDM phase boundary.
}
\end{figure}

The line of fixed points FP3 describes the phase boundary between the semimetal, and massless CDM phase for the noninteracting problem in the $\Delta_{M}-\Delta_{05}$ plane. The associated RG flow and phase diagram in $\Delta_M-\Delta_{05}$ plane, for $m=B=\alpha=\Delta_V=\Delta_{SO}=0$ are respectively shown in Fig.~\ref{fig:subfigsupp4a} and Fig.~\ref{fig:subfigsupp4b}. The mass disorder shifts the phase boundary to higher values of $\Delta_{05}$, and the diffusive phase does not have chiral symmetry.
Now linearizing about this line of fixed points we find
\begin{eqnarray}
\frac{d\delta \alpha}{dl}&=&(\frac{1}{2}+2\Delta_{M}^{\ast})\delta \alpha \Rightarrow \delta \alpha =\delta \alpha_0 e^{(1/2+2\Delta_{M}^{\ast})l}\label{eq:FP31}\\
\frac{d\delta\Delta_{V}}{dl}&=&4\Delta_{M}^{\ast}\delta\Delta_{V}+4\Delta_{M}^{\ast}\delta\Delta_{SO}\label{eq:FP32}\\
\frac{d\delta\Delta_{SO}}{dl}&=&\frac{4}{3}\Delta_{M}^{\ast}\delta\Delta_{SO}+\frac{4}{3}\Delta_{M}^{\ast}\delta\Delta_{V}\label{eq:FP33}\\
\frac{d\delta\Delta_{M}}{dl}&=&2\Delta_{M}^{\ast}(-\delta\Delta_{M}+\delta\Delta_{05}-\delta\Delta_{V}+3\Delta_{SO}+\frac{\delta \alpha}{3\pi})\label{eq:FP34}\\
\frac{d\delta\Delta_{05}}{dl}&=&2\Delta_{05}^{\ast}(\delta\Delta_{05}-\delta\Delta_{M}+\delta\Delta_{V}-3\Delta_{SO}-\frac{2\delta \alpha}{3\pi})\label{eq:FP35}
\end{eqnarray}
From Eq.~\ref{eq:FP32} and Eq.~\ref{eq:FP33}, we find $\delta\Delta_{V}-3\delta\Delta_{SO}=\delta\Delta_{V}^{0}-3\delta\Delta_{SO}^{0}$. After adding both sides of Eq.~\ref{eq:FP34} from Eq.~\ref{eq:FP35} we find
\begin{eqnarray}
\delta\Delta_{05}-\delta\Delta_{M}+2(\delta\Delta_{V}-3\delta\Delta_{SO})(\Delta_{05}^{\ast}-\Delta_{M}^{\ast})+\frac{4\delta \alpha}{3\pi}\frac{(1+3\Delta_{M}^{\ast})}{(-1+4\Delta_{M}^{\ast})}
&=&\bigg[\delta\Delta_{05}^{0}-\delta\Delta_{M}^{0}+2(\delta\Delta_{V}^{0}-3\delta\Delta_{SO}^{0})(\Delta_{05}^{\ast}-\Delta_{M}^{\ast})\nonumber \\&&+\frac{4\delta \alpha^{0}}{3\pi}\frac{(1+3\Delta_{M}^{\ast})}{(-1+4\Delta_{M}^{\ast})}\bigg]e^{l}
\label{eq:FP36}
\end{eqnarray}
is the most relevant variable if $\Delta_{M}^{\ast}<3/16$. This variable defines the phase boundary, and critical properties are non-universal. For $\Delta_{M}^{\ast}<3/16$, $\nu=1$. For $\Delta_{M}^{\ast}>3/16$, $\delta\Delta_{V}$ and $\delta\Delta_{SO}$ provide stronger perturbation with eigenvalue $16\Delta_{M}^{\ast}/3>1$. The linearized equations for finite $B$ and $m$ are
\begin{eqnarray}
\frac{dB}{dl}&=&-B\\
\frac{dm}{dl}&=&\frac{3}{2}m+\frac{1}{2}B
\end{eqnarray}
Notice that the scaling dimension of $m$ is 3/2, and this increases the stability of the insulating phases in the vicinity of FP3.

\begin{figure}[ht]
\centering
\subfigure[]{
\includegraphics[scale=0.85]{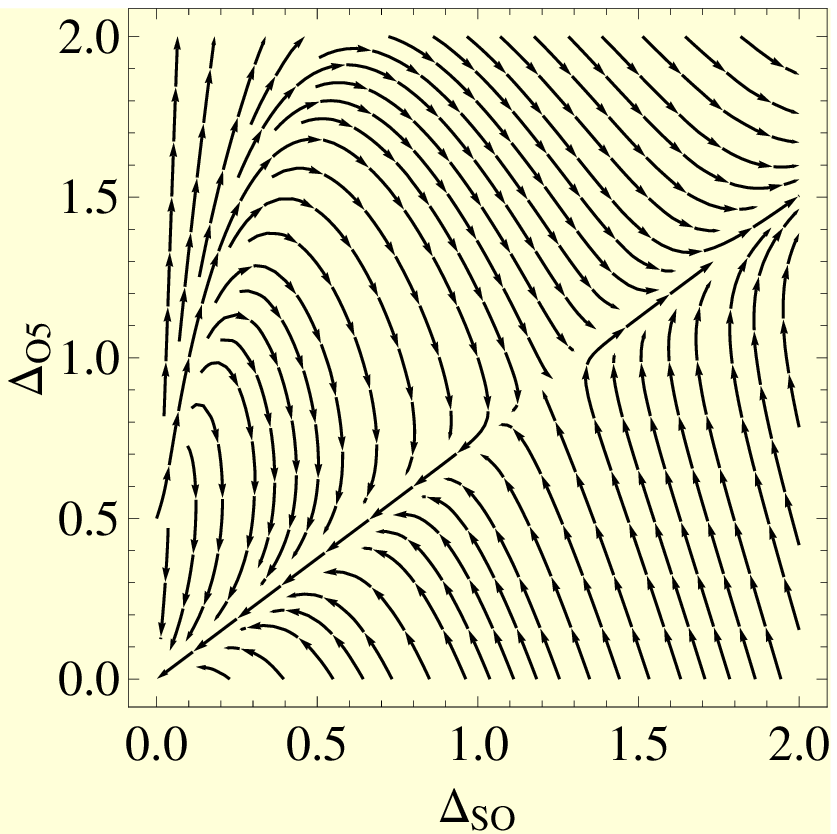}
\label{fig:subfigsupp5a}
}
\subfigure[]{
\includegraphics[scale=0.85]{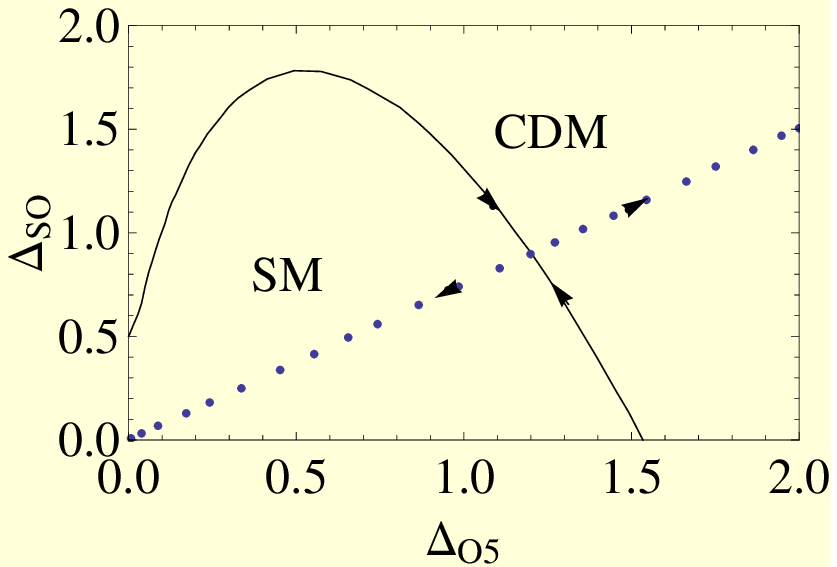}
\label{fig:subfigsupp5b}
}
\label{fig:supp5}
\caption[]{(a) The RG flow, and (b)the phase diagram in $\Delta_{SO}-\Delta_{05}$ plane for $m=B=\alpha=\Delta_V=\Delta_{M}=0$. The blue dotted line, and the phase boundary in (b) respectively correspond to the relevant and irrelevant variables at FP4.}
\end{figure}

\begin{figure}[ht]
\centering
\subfigure[]{
\includegraphics[scale=0.85]{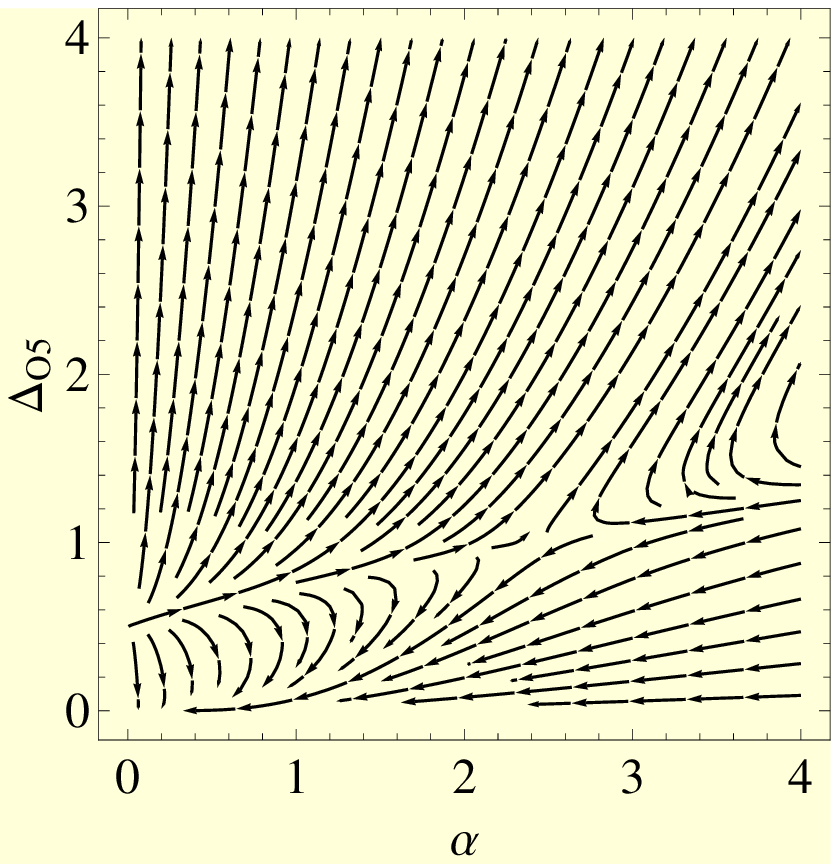}
\label{fig:subfigsupp6a}
}
\subfigure[]{
\includegraphics[scale=0.85]{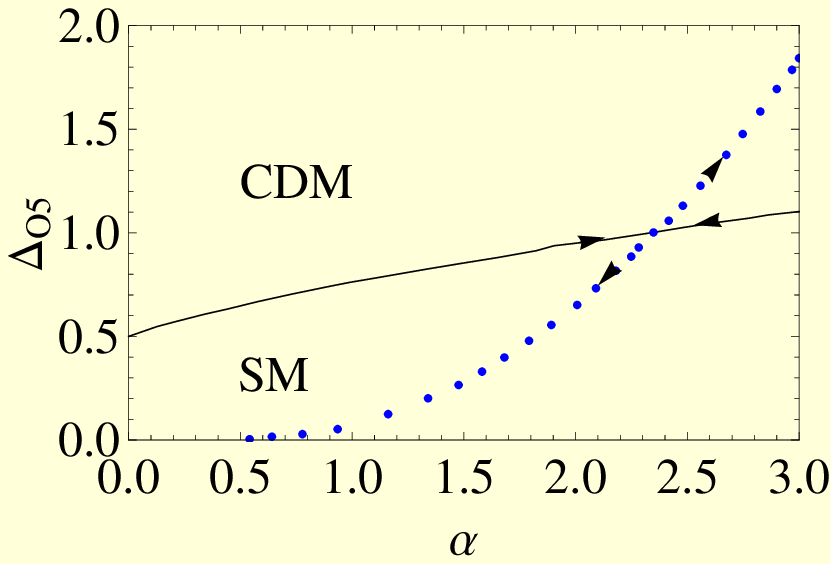}
\label{fig:subfigsupp6b}
}
\label{fig:supp6}
\caption[]{(a) The RG flow, and (b)the phase diagram in $\alpha-\Delta_{05}$ plane for $m=B=\Delta_V=\Delta_{M}=\Delta_{SO}=0$. The blue dotted line, and the phase boundary in (b) respectively correspond to the relevant and irrelevant variables at FP5.
}
\end{figure}

The fixed point FP4 is the critical point in the class of spin-orbit disorder. The associated RG flow and phase diagram in $\Delta_{SO}-\Delta_{05}$ plane, for $m=B=\alpha=\Delta_V=\Delta_{M}=0$ are respectively shown in Fig.~\ref{fig:subfigsupp5a} and Fig.~\ref{fig:subfigsupp5b}.
Linearizing about this critical point we find $\nu=1$. The irrelevant variable provides the phase boundary. This critical point is highly unstable against random potential, and mass disorder, and Coulomb interaction.
The fixed point FP5 is the only finite interaction, dirty critical point. The associated RG flow and phase diagram in $\alpha-\Delta_{05}$ plane, for $m=B=\Delta_V=\Delta_M=\Delta_{SO}=0$ are respectively shown in Fig.~\ref{fig:subfigsupp6a} and Fig.~\ref{fig:subfigsupp6b}. The Coulomb interaction shifts the phase boundary to higher values of $\Delta_{05}$.
Linearizing the flow equations we find $\delta\Delta_{V}=\delta\Delta_{V}^{0}e^{-l}$, $\delta\Delta_{M}=\delta\Delta_{M}^{0}e^{3l/2}$, and $\delta\Delta_{SO}=\delta\Delta_{SO}^{0}e^{-l}$. From the equations for $\Delta_{05}$ and $\alpha$ we find the relevant and irrelevant combinations have eigenvalues $(1\pm \sqrt{5})/2$. The irrelevant variable provides the phase boundary.

\section{II. Quantum critical scaling properties of 3+1-dimensional massless Dirac fermion} In this section we consider the quantum critical scaling properties of massless Dirac fermions when disorder is irrelevant, and for simplicity only consider the chemical potential disorder. At a finite temperature or a chemical potential the flow toward infrared limit will be truncated by the largest energy scale. Thus $\ell= \min\{\ell_T, \ell_{\mu}\}$ acts as an infra-red cutoff, where $\ell_T \sim v/T$, and $\ell_{\mu}\sim v/\mu$ are respectively the thermal de Broglie wavelength and the inter-particle separation. For $\ell_T<\ell_\mu$, we observe critical properties of the Dirac fermions, and the conventional Fermi liquid behavior is observed for $\ell_T>\ell_\mu$. For this reason we focus on the regime $\ell_T<\ell_\mu$. With decreasing temperature, the disorder strength decreases according to $\Delta_V(T) \sim \Delta_{V0} T/T_0$, where the ultra-violet scale $T_0=\hbar v \Lambda/k_B \sim 10^4 K$. In this regime there is an initial enhancement of $\alpha$ (down to a scale $T(\Delta_{V0})<T_0$), followed by the logarithmic decrease of $\alpha$ (below $T(\Delta_{V0})$). This leads to non-monotonic temperature dependent corrections to the scaling properties of non-interacting Dirac fermions.

\paragraph{\bf{Specific heat and compressibility}:} For the noninteracting clean problem, the density of states at the Fermi point vanishes quadratically, $\rho(E)\propto E^2$ and the free energy density has the power law dependence $f\sim \ell^{-(d+z)}$. The free energy density is given by
\begin{equation}
f(T,\mu)=\frac{2T^4}{\pi^2 v^3}\left[\mathrm{Li}_{4}(-e^{\mu/T})+\mathrm{Li}_{4}(-e^{-\mu/T})\right].
\end{equation}
In the limit $\mu/T\ll1$,
\begin{equation}
f(T,\mu)\approx -\frac{T^4}{v^3}\left[\frac{7\pi^2}{180}+\frac{\mu^2}{6T^2}+\frac{\mu^4}{12\pi^2T^4}\right]
\end{equation}
The specific heat follows the scaling relation $C\sim T^{d/z}$, and is given by
\begin{equation}
C=-T\frac{\partial^2f}{\partial T^2}\approx\frac{T^3}{v^3}\left[\frac{7\pi^2}{15}+\frac{\mu^2}{3T^2}\right],
\end{equation}
The compressibility also follows expected scaling relation $\kappa \sim T^{d/z-1}$ and is given by,
\begin{equation}
\kappa=-\frac{\partial^2f}{\partial \mu^2}=-\frac{2T^2}{\pi^2v^3}\left[\mathrm{Li}_{2}(-e^{\mu/T})+\mathrm{Li}_{2}(-e^{-\mu/T})\right].
\end{equation}
In the limit $\mu/T \ll 1$We find  $\kappa\approx v^{-3}(T^2/3+\mu^2/\pi^2)$. For $\mu=0$, the ratio $C/(\kappa T)=7\pi^2k_{B}^{2}/5$ is a universal number. Due to interplay of disorder and interaction, there is an initial enhancement of $C$, and $\kappa$ followed by the logarithmic suppression by a factor $(1+4\alpha_0/(3\pi)\log(T_0/T))^{-3/2}$. The non-monotonic behavior becomes pronounced if $\Delta_{V0}$ is close to the critical strength.

\paragraph{\bf{Diamagnetism}:} We first consider the clean, noninteracting problem. The magnetic field $B$ introduces another length $\ell_B=(eB)^{-1/2}$. At $T=\mu=0$, a naive application of the scaling formula  gives $f \sim (eB)^2$, and a constant diamagnetic susceptibility $\chi$. In the presence of gauge field naive scaling hypothesis becomes inapplicable. The interaction of electrons with an external gauge field requires the use of RG scheme with a proper regularization procedure to address the ultraviolet divergence of the fermionic polarization bubble. Since $d+z=4$, the problem is at upper critical dimension and one should anticipate logarithmic corrections \cite{Ghosal}. A proper analysis leads to the renormalization of charge and the gauge field, and a $(eB)^2 \log(a/\ell)$ contribution to the free energy density \cite{Salam, Blau}. Again $\ell$ has to be chosen to be the smallest of the scales chosen from $(eB)^{-1/2}$, $v/T$, $v/\mu$, and $v/m$, and accounts for the large value of $\chi$. Such logarithmic enhancement has been argued for the large diamagnetic susceptibility of $Bi$, and the narrow gap semiconductors such as $\textrm{Pb}_{1-x}\textrm{Sn}_{x}\textrm{Te}$, $\textrm{Bi}_{1-x}\textrm{Sb}_{x}$.

The Landau level spectrum of massless Dirac fermion is $E_{n,k_z}=\pm\sqrt{v^2k_{z}^{2}+E_{B}^{2}n}$; where $E_{B}=\sqrt{2}v/\ell_{B}$ is, and $n$ is the Landau level index. The degeneracy of $n=0$ level per unit area is $eB/2\pi$ and that for levels with $n>0$ is $eB/\pi$. The diamagnetic susceptibility arises from $n \neq 0$ levels, and in the following we consider the part of the free energy density $E^{'}$ that arises from $n \neq 0$ levels. This is given by
\begin{eqnarray}
E^{'}=\lim_{\epsilon \to 0}\frac{-v}{\pi^2 \ell_{B}^{4}}\left(\frac{\Lambda \ell_{B}}{\sqrt{2}}\right)^{\epsilon}\int_{-\infty}^{\infty}dx \sum_{n=1}^{\infty}\left(n+x^2\right)^{1/2-\epsilon/2}=\frac{-v\sqrt{\pi}}{\pi^2 \ell_{B}^{4}}\lim_{\epsilon \to 0}\left(\frac{\Lambda \ell_{B}}{\sqrt{2}}\right)^{\epsilon}\frac{\Gamma\left(-1+\frac{\epsilon}{2}\right)}{\Gamma\left(\frac{-1}{2}+\frac{\epsilon}{2}\right)}\zeta\left(-1+\frac{\epsilon}{2}\right)
\end{eqnarray}
The divergent part of $E^{'}$ is given by
\begin{equation}
E^{'}_{div}=\frac{-v}{\pi^2 \ell_{B}^{4}}\zeta(-1)\frac{1}{\epsilon},
\end{equation}
and it is absorbed into the vacuum energy density $B^2/2$, which leads to the field and the charge renormalizations
\begin{eqnarray}
B_{R}^{2}&=&B^2 \left(1+\frac{e^2v}{12\pi^2c^2\epsilon}\right) \\
e_{R}^{2}&=&e^2 \left(1+\frac{e^2v}{12\pi^2c^2\epsilon}\right)^{-1}
\end{eqnarray}
In the above equations $1/\epsilon=\log(\Lambda \ell_{B}/\sqrt{2})$, and we have restored the explicit dependence on the speed of light $c$. The finite part of the energy density is
\begin{eqnarray}
E^{'}_{finite}=-\frac{v}{\pi^2 \ell_{B}^{4}}\bigg[\zeta(-1)\log\left(\frac{\Lambda^2\ell_{B}^{2}}{2}\right)+ \zeta^{'}(-1)+ \zeta(-1)\left(\psi(2)-\psi(-1/2)\right)\bigg]
\end{eqnarray}
where $\psi$ is the digamma function, and $\zeta$ is the Riemann zeta function. From the finite part of the energy density we find the diamagnetic susceptibility
\begin{eqnarray}
\chi\approx-\frac{e_{R}^{2}v}{24\pi^2}\left(\log\left(\frac{B_0}{B}\right)+1.74\right), \;  \ell_{B} < \sqrt{2}\ell_{T}
\end{eqnarray}
where $B_0=\hbar \Lambda^2/(2e)\sim 10^4 T$. At a finite temperature, a similar calculation can be performed for the free energy density and in the high temperature limit we find
\begin{equation}
\chi\approx-\frac{e_{R}^{2}v}{24\pi^2}\left(\log\left(\frac{T_0}{T}\right)+1.74\right), \;  \ell_{B} > \sqrt{2}\ell_{T}
\end{equation}
The departure from the noninteracting formula follows from the renormalization of $e^2v \propto \alpha v^2$. When $\alpha$ decreases logarithmically, we can set $z \approx 1$, and find $e^2v=\mathrm{constant}$. Therefore Coulomb interaction does not modify the scaling behavior of $\chi$. The disorder only causes a small suppression of noninteracting value of $\chi$.

\paragraph{\bf{Dynamic conductivity}:} Consider again the noninteracting, clean limit. Since conductivity $\sigma \sim \ell^{-(d-2)}$, and $z=1$, we can write
$\sigma(\omega,T)=e^2T/(2\pi v)\Phi(\omega/T,\mu/T)$, where the scaling function
\begin{eqnarray}
\Phi(x,y)=1/36[(8\pi^2+24y^2)\delta(x) +3x\{\tanh(x/4+y/2)+\tanh(x/4-y/2) \}].
\end{eqnarray}
Again we will focus on $\mu=0$, or $(\omega,T)\gg \mu$ limit. The inelastic scattering rate $\tau_{in}^{-1}\sim \alpha^2 T$, is larger than the elastic scattering rate due to disorder, and the conductivity will be mainly governed by the interaction effects. If $\omega \gg \alpha^2T$, the leading order answer for the dynamic conductivity follows from the noninteracting formula,
\begin{equation}
\sigma(\omega,T)\approx \frac{e^2\omega}{12\pi v}\tanh(\omega/4T)
\end{equation}
As $\sigma(\omega,T) \propto \alpha$, there is an initial enhancement, followed by a logarithmic suppression by the factor $\alpha$$\left[1+\frac{4\alpha_0}{3\pi}\log\left(\frac{T_0}{\omega}\right)\right]^{-1}$.

In the opposite limit $\omega \ll \alpha^2 T$ the collision processes in the particle-hole plasma due to Coulomb interaction governs the conductivity \cite{Fritz, Arnold}. To obtain concrete answer we have performed a calculation using the quantum Boltzman equation
\begin{equation}
(\partial_t+e\mathbf{E}\cdot \nabla_{\mathbf{p}})f_a(\mathbf{p},t)=-C[f_a](\mathbf{p},t)
\end{equation}
within leading log approximation \cite{Fritz, Arnold}. In the above equation $\mathbf{E}$ is the external electric field and $f_a(\mathbf{p},t)$ is the fermion distribution function, and $a$ is the collective label for particles and holes, and also the chiralities. The definition of the collision operator $C[f_a](\mathbf{p},t)$ involves the square of the amplitudes of the two particle scattering processes (particle-particle, hole-hole and particle-hole), and a combination of Fermi functions. In order to solve the Boltzman equation we introduce the ansatz
\begin{equation}
\frac{f_a(\mathbf{p},\omega)}{f^{0}_{a}(\mathbf{p})}=2\pi \delta(\omega)+(1-f^{0}_{a}(\mathbf{p}))e\mathbf{E}(\omega)\cdot\frac{\hat{p}}{T^2}\chi_a(\mathbf{p},\omega)
\end{equation}
where $f^{0}_{a}$ is equilibrium distribution function, and convert the linearized Boltzman equation into a varitaional problem for
$\chi_a(\mathbf{p},\omega)$ which are functions of dimensionless variables $|\mathbf{p}|/T$ and $\omega/T$. In the particle-hole symmetric case $\mu=0$, $\chi_+(\mathbf{p},\omega)=-\chi_-(\mathbf{p},\omega)$, and particles and holes equally contribute to the transport. At the end we extremize the functional $Q[\chi]$, given by
\begin{eqnarray}
\frac{Q[\chi]}{T^2}=\int_{0}^{\infty} dp f^{0}(p)(1-f^0(p))\left[\frac{e^4}{144\pi^3} \left\{\left(p\chi^{\prime}(p,\omega)\right)^2+\frac{2}{p^2}\chi^{2}(p,\omega)\right\}-\frac{2}{T^2}\left\{\chi(\mathbf{p},\omega)+\frac{i\omega}{2T}\chi^{2}(\mathbf{p},\omega)\right\}\right]
\end{eqnarray}

Now choosing a single parameter ansatz $\chi(p,\omega)=(p/T)^{n}g(\omega/T)$, we find the extremum occurs for $n\sim 0.896$, and
\begin{equation}
\sigma(\omega,T)=\frac{30.46T}{\alpha \log(1/\alpha)}\left[1-\frac{i\omega}{T}\times\frac{26.67}{\alpha^2 \log(1/\alpha)}\right]^{-1}
\end{equation}
One popular choice $n=1$ \cite{Baym}, only accounts for particle-hole scattering. The proximity of our $n\sim 0.896$ to 1, suggests that the contribution from  particle-particle collisions is small. Only for $\mu>T$, the like-particle collisions will be dominant and lead to conventional Fermi liquid result $\tau^{-1}\sim \alpha^2 T^2/\mu$, and $\sigma_0 \sim \mu^3/(\alpha T^2)$. From the expression for dynamic conductivity we can see the existence of a Drude peak, and $T$-linear dc conductivity. The renormalization of $\alpha$ will now cause an initial suppression, followed by a logarithmic enhancement of the dc conductivity.


\begin{thebibliography}{}
\bibitem{Nimtz}R. Dornhaus, G. Nimtz, and B. Schlicht, {\it Narrow-Gap Semicounductors}, (Springer-Verlag, 1983).

\bibitem{FuKane1}L. Fu, and C. L. Kane, Phys. Rev. B \textbf{76}, 045302 (2007)

\bibitem{QiZhang1}X. L. Qi, T. L. Hughes, and S. C. Zhang, Phys. Rev. B \textbf{78}, 195424 (2008)

\bibitem{Zahid1}M. Z. Hasan, and C. L. Kane, Rev.Mod.Phys. \textbf{82}, 3045 (2010).

\bibitem{QiZhang2}X. L. Qi, and S. C. Zhang, arXiv:1008.2026v1

\bibitem{Zhang}H. Zhang {\it et al.}, Nat. Phys., \textbf{5}, 438 (2009).

\bibitem{Liu}C. X. Liu {\it et al.}, arXiv:1005.1682v1

\bibitem{Abrikosov}A. A. Abrikosov and S. D. Beneslavskii, Sov. Phys. JETP \textbf{32}, 699 (1971).

\bibitem{Fradkin}E. Fradkin, Phys. Rev. B \textbf{33}, 3263 (1986).

\bibitem{Shindou}R. Shindou and S. Murakami, Phys. Rev. B \textbf{79}, 045321 (2009).

\bibitem{Guo}H. M. Guo {\it et al.}, Phys. Rev. Lett. \textbf{105}, 216601 (2010).

\bibitem{Ryu}A. P. Schnyder {\it et al.},  Phys. Rev. B \textbf{78}, 195125 (2008).

\bibitem{Ostrovsky}P. M. Ostrovsky, I. V. Gornyi, and A. D. Mirlin, Phys. Rev. Lett. \textbf{105}, 036803 (2010).

\bibitem{Supp}EPAPS Document No. X-XXX-XXXXXX-XXX-XXXXXX

\bibitem{Foster} Matthew S. Foster and Igor L. Aleiner, Phys. Rev. B \textbf{77}, 195413 (2008).

\bibitem{Schmalian} D. E. Sheehy and J. Schmalian, Phys. Rev. Lett. \textbf{99}, 226803 (2007).

\bibitem{Loring}M. B. Hastings, and T. A. Loring, arxiv:1012.1019.

\bibitem{Belitz}D. Belitz, and T. R. Kirkpatrick, Rev. Mod. Phys. \textbf{66}, 261 (1994).

\end{thebibliography}

\begin{thebibliography}{19}

\bibitem{Ghosal} A. Ghosal, P. Goswami, and S. Chakravarty, Phys. Rev. B \textbf{75}, 115123 (2007).

\bibitem{Salam} A. Salam and J. Strathdee, Nucl. Phys. B \textbf{90}, 203 (1975).

\bibitem{Blau} S. K. Blau, M. Visser, and A. Wiff, Int. J. Mod. Phys. \textbf{6}, 5409 (1991).

\bibitem{Fritz} L. Fritz {\it et al.}, Phys. Rev. B \textbf{78}, 085416 (2008).

\bibitem{Arnold} P. Arnold {\it et al.}, J. High Energy Phys. \textbf{11}, 001 (2000).

\bibitem{Baym} G. Baym and H. Heiselberg, Phys. Rev. D \textbf{56}, 5254 (1997).

\end{thebibliography}
\end{document}